\documentclass[prd,nofootinbib,showpacs,preprint]{revtex4}
\usepackage{graphicx}
\usepackage{epsfig}
\usepackage{amsmath}
\usepackage{amsfonts}
\usepackage{amssymb}
\newcommand{\beqa}{\begin{eqnarray}} 
\newcommand{\eeqa}{\end{eqnarray}} 
\newcommand{\beq}{\begin{equation}} 
\newcommand{\eeq}{\end{equation}}

\newcommand{\nn}{\nonumber}
\newcommand{\be}{\begin{equation}}
\newcommand{\ee}{\end{equation}}
\newcommand{\bea}{\begin{eqnarray}}
\newcommand{\eea}{\end{eqnarray}}

\begin{document}
\title{Leptoquark effects on $b \to s \nu \bar{\nu}$ and $B \to K l^+ l^-$ decay  processes}
\author{Suchismita Sahoo and Rukmani Mohanta }
\affiliation{\,School of physics, University of Hyderabad, 
              Hyderabad - 500046, India  }      
\begin{abstract}
We study the rare semileptonic decays of $B$ mesons induced by $b \to s \nu \bar{\nu}$ as well as $b \to s l^+ l^-$ 
transitions in the scalar leptoquark model where the leptoquarks transform as $(3,2,7/6)$ and $(3,2,1/6)$
under the standard model gauge group. The leptoquark parameter space is constrained using the most recent experimental results  on 
${\rm Br}(B_s \rightarrow \mu^+ \mu^-)$ and ${\rm Br}(B_d \to X_s \mu^+ \mu^-)$ processes. Considering  only  the baryon number conserving 
leptoquark interactions, we estimate the branching ratios for the exclusive $\bar{B} \rightarrow \bar{K}^{(*)} \nu \bar{\nu}$  and 
inclusive $B \rightarrow X_s \nu \bar{\nu}$ decay processes by using the constraint parameters. 
We also obtain the low recoil (large lepton invariant mass, i.e., $q^2 \sim m^2_b$) predictions for the angular distribution of  
$\bar B \rightarrow \bar{K} l^+ l^-$ process and several other observables including the flat term and lepton
flavour  non-universality  factor in this model.
\end{abstract}
\maketitle
\section{introduction}

It is well-known that  the study of $B$ physics plays an important role to critically test the standard model (SM) predictions and to look for 
possible signature of new physics beyond it. In particular, the rare decays of $B$ mesons which are mediated by flavour changing neutral current (FCNC) 
transitions are well-suited for searching the effects of possible new interactions beyond the SM. This is due to the fact that the FCNC transitions 
$b \to s,d$ are highly suppressed in the SM as  they occur only at one-loop level and hence, they are very sensitive to new physics. Recently the decay modes $B \to K^{(*)} l^+ l^-$, which
are mediated by the quark level transition $b \to s l^+ l^-$ have attracted a lot of attention, as several anomalies  at the 
level of few sigma are observed in the LHCb experiment \cite{lhcb1,lhcb2,lhcb7}. 
Furthermore, the  deviation in the ratio of rates of $B \to K \mu \mu$ over  $B \to K e e$ ($R_K$) is a hint of violation of lepton universality 
\cite{lhcb3}. This in turn requires the careful analyses of the angular observables for these processes both in the low and high $q^2$ regime.

Recently various $B$ physics experiments such as BaBar, Belle, CDF and LHCb have provided data on the angular distributions of 
$B \to K^* l^+ l^- $ and $B \to K l^+ l^- $ decay processes both in the low and the large recoil region except the intermediate region 
around $q^2 \sim m^2_{J/\psi}$ and $m^2_{\psi^\prime}$. The intermediate region is dominated by the pronounced charmonium resonance background 
induced by the decays $B \to K (\bar{c}c) \to K l^+ l^- $, where $\bar{c}c = J/\psi, \psi^\prime$. Using QCD factorization method
the physical observables in the high recoil region can be calculated and  the angular distribution of $\bar{B} \to \bar{K} l^+ l^-$ at 
low recoil can be computed using simultaneous heavy quark effective theory and operator product expansions in $1/Q$, 
with $Q = (m_b, \sqrt{q^2})$ \textit{i.e.} $\sqrt{q^2}$ is of the order of the $b$-quark mass \cite{grinstein2, beylich}. 
In this work, we are interested to study the decay process  $B \to K l l $  in the region of low hadronic recoil \textit{i.e.} above the $\psi^\prime$ peak
in the scalar leptoquark (LQ) model.
We have studied the  $B \to K \mu^+ \mu^-$ in the large recoil limit in Ref. \cite{mohanta2} and  found that the various anomalies associated 
with the isospin asymmetry  parameter and the lepton flavour non-universality factor ($R_K$) for this process
can be  explained in this model.

Similarly the rare  semileptonic decays of $B$ mesons    with $\nu \bar{\nu}$ pair in the final state, i.e., $B\to K^{(*)} \nu \bar{\nu} $ are also significantly suppressed  
in the SM  and their long distance contributions are generally subleading. These decays are theoretically very clean due to the absence of photonic penguin 
contributions and strong suppression of light quarks. The experimental measurement of the inclusive decay rate  probably be un-achievable 
 due to the missing neutrinos, however, the exclusive channels like $B \to K^* \nu \bar{\nu}$ and  $B \to K \nu \bar{\nu}$ are 
more promising as far as the  measurement of branching ratios and other related observables are concerned. Theoretically, study of these decays requires calculation of 
relevant form factors by non-perturbative  methods.

In recent times, there are many interesting papers which are contemplated to explain  the anomalies associated with the  $b \to s l^+ l^-$ processes, 
observed at LHCb  experiment \cite{lhcb1, lhcb2, lhcb7, lhcb3},  both in the context of various new physics models as well as  in model 
independent ways \cite{matias1, jager, huber, beaujean}. In this paper, we intend to  study the effect of scalar  leptoquarks 
transform as $(3,2,7/6)$ and $(3,2,1/6)$ under the standard model gauge group,
on the branching ratio as well as on other  asymmetry parameters in the low-recoil region of  $B \to K l^+ l^-$  process.
We also consider the  processes $B \to K^{(*)} \nu \bar{\nu}$ and $B \to X_s \nu \bar{\nu}$
involving the quark level transitions $b \to s \nu \bar{\nu}$ in the full physical regime. It is well-known that 
leptoquarks are scalar or vector color triplet 
bosonic particles which make leptons couple directly to quarks and vice versa and carry both lepton as well as baryon quantum numbers and fractional electric 
charge. Leptoquarks can be included in the low energy theory as a relic of a more fundamental theory at some high energy scale in the 
extended SM \cite{georgi}, such as grand unified theories \cite{georgi, georgi2}, Pati-Salam models, models of extended technicolor 
\cite{schrempp} and composite models \cite{kaplan}. Leptoquarks are classified by their fermion number ($F=3B+L$), spin and charge. 
 Usually they have a mass near the 
unification scale to avoid  rapid proton decay, even so leptoquarks may exist at a mass accessible to present collider, if baryon and lepton 
numbers would conserve separately. The leptoquark properties and the additional new physics contribution to the SM have been very well 
studied in the literature \cite{davidson,arnold,kosnik,mohanta2,mohanta1,leptoquark}.

The plan of the paper is follows. In section II  we present the effective Hamiltonian responsible for $b \rightarrow s l^+ l^-$ processes.
 We also discuss the new physics contributions due to the exchange of scalar leptoquarks. In section III we discuss the 
constraints on leptoquark parameter space by using the recently measured branching ratios of the rare decay modes $B_s \rightarrow \mu^+ \mu^- $
and $B_d \to X_s \mu^+ \mu^-$. 
The branching ratio, the flat term and the lepton non-universality factor $(R_K)$ for the decay mode $\bar{B} \to \bar{K} l^+ l^- $, 
where $l = e,\mu, \tau$ at low recoil 
limit are computed in section IV. In section V we work out the branching ratio of $\bar{B} \to \bar{K} \nu \bar{\nu}$ process in the full kinematically  
accessible  physical region. The branching ratio, polarization and other asymmetries in $\bar{B} \to \bar{K}^* \nu \bar{\nu}$ process  have been 
computed in section VI. The inclusive decay process $B \to X_s \nu \bar{\nu}$ is discussed in section VII and section 
VIII contains the summary and conclusion.

\section{ The Effective Hamiltonian for $b \to s l^+ l^- $ process } 
The effective Hamiltonian  describing the processes induced by the FCNC $b \rightarrow sl^+ l^-$ transitions  is given by \cite{buras1}
\bea
{\cal H}_{eff} &=& - \frac{ 4 G_F}{\sqrt 2} V_{tb} V_{ts}^* \Bigg[\sum_{i=1}^6 C_i(\mu) O_i +\sum_{i=7,9,10} \left ( C_i(\mu) O_i
+ C_i'(\mu) O_i' \right )
\Bigg]\;,\label{ham}
\eea
which consists of the tree level current-current operators ($O_{1,2}$),
QCD penguin operators ($O_{3-6}$) alongwith the magnetic $O_7^{(\prime)}$ and semileptonic electroweak penguin operators $O_{9,10}^{(\prime)}$.
The magnetic and electroweak penguin operators can be expressed as 
\bea
O_7^{(\prime)} &=&\frac{e}{16 \pi^2} \Big(\bar s \sigma_{\mu \nu}
(m_s P_{L(R)} + m_b P_{R(L)} ) b\Big) F^{\mu \nu} \nn\\
O_9^{(\prime)}&=& \frac{\alpha}{4 \pi} (\bar s \gamma^\mu P_{L(R)} b)(\bar l \gamma_\mu l)\;,~~~~~~~ O_{10}^{(\prime)}= \frac{\alpha}{4 \pi} (\bar s \gamma^\mu 
P_{L(R)} b)(\bar l \gamma_\mu \gamma_5 l)\;.
\eea
It should be noted that the primed operators are absent in the SM.
The  values of Wilson coefficients $C_{i=1, \cdots, 10}$,  which are  evaluated 
in the next-to-next leading order at the  renormalization scale $\mu = m_b$ are taken from  \cite{kohda}. 
Here $ V_{qq^\prime}$ denotes the CKM matrix element, $G_F$ is the Fermi constant, $\alpha$ is the fine-structure constant and 
$P_{L,R} = (1\mp \gamma_5)/2$ are the chiral projectors. Due to the negligible contribution of the CKM-suppressed factor $V_{ub} V_{us}^*$, 
there is no CP violation in the decay amplitude in the SM. These processes will receive additional contributions  due to
the exchange of scalar leptoquarks. In particular there will be new contributions to the electroweak penguin operators 
$O_9$ and $O_{10}$ as well their right-handed counterparts $O_9'$ and $O_{10}'$.
In the following subsection we will present these  
additional contributions to the SM effective Hamiltonian due to the exchange of such leptoquarks.

\subsection{Scalar LQ Contributions to $b \to s l^+ l^- $ effective Hamiltonian}

There are  ten different types of leptoquarks  under the $SU(3)\times SU(2)\times U(1)$ gauge group \cite{ref45}, half of them  have scalar nature and other
halves have vector nature under the Lorentz transformation. 
The scalar leptoquarks have spin zero and could  potentially contribute to
the quark level transition $b \to s l^+ l^-$.  Here we would like to consider the minimal 
renormalizable scalar leptoquark model \cite{arnold}, containing one single additional representation of $SU(3)\times SU(2)\times U(1)$, 
which does not allow proton decay. There are only two such models  with representations under the SM gauge group as
$\Delta^{(7/6)}\equiv (3,2,7/6)$ and $\Delta^{(1/6)}\equiv(3,2,1/6)$ \cite{arnold}, which  have sizeable Yukawa couplings to matter fields.  
These scalar leptoquarks do not have baryon number violation in the perturbation theory 
and could be light enough  to be accessible in accelerator searches.
The interaction Lagrangian of the scalar leptoquark $\Delta^{(7/6)}$ with the fermion bilinear is given as \cite{kosnik}
\begin{equation}
\mathcal{L}^{(7/6)} = g_R \bar{Q}_L\Delta^{(7/6)} l_R  + h.c.,
\end{equation}
where $Q_L$ is the left handed quark doublet and $l_R$ is the right-handed charged lepton singlet.
After performing the Fierz transformation and comparing with the SM effective Hamiltonian (\ref{ham}), 
one can obtain the new Wilson coefficients as discussed in Ref. \cite{kosnik}
\begin{equation}
C_9^{NP} = C_{10}^{NP} = -\frac{\pi}{2\sqrt{2}G_f \alpha V_{tb}V_{ts}^*}\frac{(g_R)_{sl}(g_R)_{bl}^* }{M_{\Delta^{(7/6)}}^2}\;.
\end{equation}
Similarly, the Lagrangian for the coupling of scalar leptoquark $\Delta^{(1/6)} $ to the SM  fermions is given by
\begin{equation}
\mathcal{L}^{(1/6)} = g_L \bar{d_R}\tilde{\Delta}^{(1/6)\dagger} L + h.c., \hspace{0.5cm} {\rm with} \hspace{0.5cm} \tilde{\Delta} \equiv i\tau_2 \Delta^*,
\end{equation}
where $\tau_2$ is the  Pauli matrix and consists of operators with right-handed quark currents. Proceeding like the previous case
one can obtain the new Wilson coefficients as
\begin{equation}
C_9^{\prime NP} = -C_{10}^{\prime NP} = \frac{\pi}{2\sqrt{2}G_f \alpha V_{tb}V_{ts}^*}\frac{(g_L)_{sl}(g_L)_{bl}^*}{M_{\Delta^{(1/6)}}^2},
\end{equation}
which are associated with the right-handed semileptonic electroweak penguin operators $O_9'$ and $O_{10}'$. 

\section{Constraint on the LQ parameters}

After having the idea of  possible scalar leptoquark contributions to the $b \to s ll$ processes we now proceed to constraint the LQ couplings using the 
theoretical \cite{bobeth1} and experimental branching ratio \cite{cms, lhcb5, lhcb6} of $B_s \to \mu^+ \mu^-$ process. This 
process is mediated by $b \to s \mu \mu $ transition and hence well-suited for constraining the LQ parameter space. In the SM the branching ratio
for this process depends only on the Wilson coefficient $C_{10}$. However, in the scalar LQ model there will be additional contributions
due to the leptoquark exchange which are characterized by the new Wilson coefficients $C_{10}^{NP}$
and $C_{10}^{'NP}$ depending on the nature of the LQs. Thus, in this model the branching ratio has the form \cite{mohanta1, mohanta2}
\bea
{\rm Br}(B_s \to \mu^+ \mu^-) = \frac{G_F^2}{16 \pi^3} \tau_{B_s} \alpha^2 f_{B_s}^2 M_{B_s} m_{\mu}^2 |V_{tb} V_{ts}^*|^2
\left |C_{10}^{SM}+C_{10}^{NP}-C_{10}^{'NP}\right |^2 \sqrt{1- \frac{4 m_\mu^2}{M_{B_s}^2}}\;,
\eea
which can be expressed  as   
\bea
{\rm Br}(B_s \to \mu^+ \mu^-)={\rm Br}^{SM}\left | 1+ \frac{C_{10}^{NP} - C_{10}^{' NP}}{C_{10}^{SM}} \right |^2
\equiv {\rm Br}^{SM}\left | 1+ r e^{i \phi^{ NP}} \right |^2\;,
\eea
where ${\rm Br}^{SM}$ is the SM branching ratio and  we define the parameters $r$ and $\phi^{NP}$ as
\be
r e^{i \phi^{NP}}=\frac{C_{10}^{NP} - C_{10}^{' NP}}{C_{10}^{SM}}\;.
\ee
Now comparing the SM theoretical prediction of ${\rm Br}(B_s \to \mu \mu)$ \cite{bobeth1}
\bea
{\rm Br}(B_s \to \mu^+ \mu^-)|_{\rm SM}&=&\left (3.65 \pm 0.23 \right ) \times 10^{-9},
\eea
with the corresponding experimental value 
\bea
{\rm Br}(B_s \to \mu^+ \mu^-)=\left (2.9 \pm 0.7 \right ) \times 10^{-9},
\eea
one can obtain
the constraint on the new physics  parameters $r$ and $\phi^{NP}$. The constraint on the leptoquark parameter space has been extracted in 
\cite{mohanta2, mohanta1} from this process, therefore,  here we will 
simply quote the results.
The allowed parameter space
in $r-\phi^{NP}$ plane which is compatible with the $1\sigma$ range of
the experimental data is 
$0\leq r \leq 0.1 $ for the entire range of $\phi^{NP}$, i.e.,
\bea
 0\leq r \leq 0.1\;, ~~~~{\rm for}~~~~0 \leq \phi^{NP} \leq 2 \pi \;.
 \eea
However, in this analysis we will use relatively mild constraint, consistent with both measurement of 
${\rm Br}(B_s \to \mu^+ \mu^-)$ and ${\rm Br}(\bar B_d^0 \to X_s \mu^+ \mu^-)$ \cite{mohanta2} as
 \bea
 0\leq r \leq 0.35\;, ~~~~{\rm with}~~~~\pi/2 \leq \phi^{NP} \leq 3 \pi/2\;.\label{r-bound1}
 \eea
 It should be noted that the use of this limited range of CP phase, i.e.,  ($\pi/2 \leq \phi^{NP} \leq 3 \pi/2$) 
is an assumption to have a relatively larger value of $r$.  These bounds can be
  translated to obtain the bounds for the leptoquark couplings  as
 \bea
 0 \leq \frac{|(g_R)_{s \mu} (g_R)_{b \mu}^*|}{M_{\Delta}^2} \leq 5 \times 10^{-9} ~ {\rm GeV}^{-2}~~~~{\rm for}~~~~\pi/2 \leq \phi^{NP} \leq 3 \pi/2\;.
 \eea
After obtaining the bounds on leptoquark couplings, we now proceed to study the decay processes $B \to K ll$ and $B \to K^{(*)}(X_s) \nu \bar \nu$
and the associated observables in the following sections. 

\section{$\bar{B} \rightarrow \bar{K} l^+ l^-$ process in the low-recoil limit}

The transition amplitude for the $B \to K l^+ l^-$ decay  process can be obtained using the effective Hamiltonian presented in Eq. (\ref{ham}). 
The matrix elements of the various hadronic currents between
the initial $B$ meson and the final $K$ meson can be parameterized in terms of the
form factors $f_0$, $f_T$ and $f_+$ as \cite{bobeth2} 
\begin{equation}
 \langle\bar{K}\left(k\right)|\bar{s}\gamma^\mu b|\bar{B}\left(p\right)\rangle = f_+\left(q^2\right) \left(p+k\right)^\mu + \left[f_0\left(q^2\right) - f_+\left(q^2\right)\right]\frac{m^2_B - m^2_K}{q^2}q^\mu ,
 \end{equation}
 \begin{equation}
 \langle\bar{K}\left(k\right)|\bar{s} \sigma^{\mu\nu} b|\bar{B}\left(p\right)\rangle = i\frac{f_T\left(q^2\right)}{m_B + m_K} \left[\left(p+k\right)^\mu q^\nu - q^\mu \left(p+k\right)^\nu \right] ,\hspace{2.5cm}
 \end{equation}
where $p, k $ are the four-momentum of the $B$-meson and Kaon respectively and $q = p-k$ is the four-momentum transferred to the dilepton system. 
Furthermore, using the QCD operator identity \cite{bobeth4, grinstein, grinstein2},
\begin{equation}
i\partial^\nu \left(\bar{s}i\sigma_{\mu \nu}b\right) = -m_b \left(\bar{s}\gamma_\mu b \right) + i\partial_\mu \left(\bar{s}b\right) 
- 2\left(\bar{s}i\overleftarrow{D}_\mu b\right),
\end{equation}
 an improved Isgur-Wise relation between $f_T$ and $f_+$ can be obtained as
 \begin{equation}
 f_T\left(q^2 ,\mu\right) = \frac{m_B\left(m_B + m_K\right)}{q^2}\kappa\left(\mu\right)f_+\left(q^2\right) + \mathcal{O}\left(\frac{\Lambda}{m_b}\right),
\label{ff}
\hspace{3cm}
 \end{equation}
where strange quark mass has been neglected. 
Thus, one can obtain the amplitude for the $\bar{B} \rightarrow \bar{K} l^+ l^-$ process in low recoil limit  \cite{bobeth2, bobeth3}, after applying 
form factor relation (\ref{ff}) as
\begin{equation}
{\cal A} \left( \bar{B} \rightarrow \bar{K} l^+ l^- \right) = i \frac{G_F \alpha}{\sqrt{2}\pi}V_{tb}V^*_{ts}f_+(q^2)[F_V p^\mu (\bar{l}\gamma_\mu l)+F_A p^\mu (\bar{l}\gamma_\mu \gamma_5 l)+F_P (\bar{l}\gamma_5 l)],
\end{equation}
where
\begin{equation}
\begin{split}
&F_A = C_{10}^{tot} ,\hspace{3.5cm} F_V = C_9^{tot}+\kappa \frac{2m_b m_B}{q^2}C_7^{eff} , \\
& F_P =-m_l \left[1+\frac{m^2_B-m^2_K}{q^2} \left(1-\frac{f_0}{f_+}\right)\right] C_{10}^{tot}.
\end{split}\label{ff1}
\end{equation}
In Eqn. (\ref{ff1}), $C_9^{tot} = C_9^{eff}+C_9^{NP}+C_9^{'NP}$ and $C_{10}^{tot} = C_{10}^{SM}+C_{10}^{NP}-C_{10}^{'NP}$, 
where $C_9^{(\prime)NP}$ and $C_{10}^{(\prime)NP}$ are the 
new contributions to the Wilson coefficients arising due to the exchange of leptoquarks and 
the effective Wilson coefficients $C_{7,9}^{eff}$  are given in  Ref. \cite{bobeth5}.
The corresponding differential  decay distributions is given by
\begin{equation}
\frac{d^2\Gamma_l\left[\bar{B} \rightarrow \bar{K} l^+ l^-\right]}{dq^2 d\cos{\theta}_l} = a_l\left(q^2\right)+c_l \left(q^2\right)\cos^2\theta_l\;,
\label{brK}
\end{equation}
where $\theta_l$ is the angle between the directions of $\bar{B}$ meson and the $l^-$, in the dilepton rest frame.
The expressions for the  $q^2$ dependent parameters $a_l$, $c_l$ are presented in Appendix A.
Thus, the decay rate for the process $\bar{B} \rightarrow \bar{K} l^+ l^-$  can be written as
\begin{equation}
\Gamma_l = 2\int_{q^2_{min}}^{q^2_{max}}dq^2 \left(a_l + \frac{1}{3}c_l\right) .\hspace{4cm}
\end{equation}
Another useful observable known as the flat term is defined as
\begin{equation}
F^l_H = \frac{2}{\Gamma_l}\int_{q^2_{min}}^{q^2_{max}}dq^2 \left(a_l + c_l\right) ,\hspace{4cm}
\end{equation}
where  the hadronic uncertainties are reduced due to cancellation between the numerator and denominator. 
It should be noted that the lepton mass suppression of $(a_l + c_l)$ follows  as $(F^{l}_H)^{SM} \propto m^2_l $, 
hence, it vanishes in the limit $m_l \rightarrow 0$.

After obtaining the expressions for branching ratio and the observable $F^l_H$, we now proceed for numerical   estimation for $B \to K l^+ l^-$ 
process in the low recoil region. In our analysis  we use  the following parametrization for the  $q^2$ dependence of form factors $f_i$ ($i=+,T,0)$ as 
 \cite{bobeth2, mannel}
 \begin{equation}
 f_i\left(s\right) = \frac{f_i\left(0\right)}{1-s/m^2_{res,i}} \Bigg[ 1+b^i_1\left(z\left(s\right)-z\left(0\right)+\frac{1}{2}\left(z\left(s\right)^2 - z\left(0\right)^2\right)\right)\Bigg],
\end{equation} 
where we have used the notation $q^2 \equiv s$. The $z(s)$ functions are given as
\begin{equation*}
z\left(s\right) = \frac{\sqrt{\tau_+ - s}-\sqrt{\tau_+ - \tau_0}}{\sqrt{\tau_+ - s}+\sqrt{\tau_+ - \tau_0}} ,
\hspace{1cm} \tau_0 = \sqrt{\tau_+}\left(\sqrt{\tau_+} - \sqrt{\tau_+ - \tau_-}\right) ,\hspace{1cm} \tau_\pm = \left(m_B \pm m_K\right)^2 .
\end{equation*}
The values of $f_i(0)$ and $b_1^i$ are taken from \cite{bobeth2}.

For numerical evaluation, we have used the particle masses and the lifetimes of $B$ meson from \cite{pdg}. 
For the CKM matrix elements, we have used the Wolfenstein parametrization with values $A = 0.814^{+0.023}_{-0.024}$, 
$\lambda = 0.22537 \pm 0.00061$, $\bar{\rho} = 0.117 \pm 0.021$ and 
$\bar{\eta} = 0.353 \pm 0.013$ and the fine structure coupling constant $\alpha = 1/137$.
With these input parameters, the differential branching ratios for $\bar{B}_d^0 \rightarrow \bar{K}^0 e^+ e^-$ (left panel), 
$\bar{B}_d^0 \rightarrow \bar{K}^0 \mu^+ \mu^-$ (right panel)  
and $\bar{B}_d^0 \rightarrow \bar{K}^0 \tau^+ \tau^-$  (lower panel) processes with respect to high $q^2$,  both in  the SM and in the  leptoquark model are 
shown  in Fig. 1 for $\Delta^{(7/6)}$ leptoquark and in Fig. 2 for $\Delta^{(1/6)}$.  The grey bands in these plots correspond to the uncertainties arising in the SM due to
the uncertainties  associated with the CKM matrix elements and the hadronic form factors. The green bands correspond to the LQ contributions. For 
$B \to K \mu \mu$ process, we vary the values of the leptoquark couplings as given in Eq. (14) and for $B \to K ee$ and $B \to K \tau \tau$ 
processes we use the limits on the LQ couplings 
extracted from $B_d \to X_s e^+ e^-$ and $B_s \to \tau^+ \tau^-$ processes \cite{mohanta2} as
 \bea
 0 \leq \frac{|(g_R)_{s e} (g_R)_{b e}^*|}{M_{\Delta}^2} \leq 1.0 \times 10^{-8} ~ {\rm GeV}^{-2}\;,
 \eea
and
 \bea
 0 \leq \frac{|(g_R)_{s \tau} (g_R)_{b \tau}^*|}{M_{\Delta}^2} \leq 1.2 \times 10^{-8} ~ {\rm GeV}^{-2}\;.
 \eea
Since the leptoquark couplings are more tightly constrained in $b \to s \mu \mu$ transitions, the deviations of the branching ratios
in the LQ model from the corresponding  SM values
are found to be small. For $B \to K ee$ and $B \to K \tau \tau$ these deviations are found to be significantly large.  
The bin-wise  experimental values are shown in black  in $B \to K \mu \mu$ process. From these figures it can be seen that 
the observed experimental data can be explained in the scalar LQ model but the deviation from the SM branching ratios are more in the 
$\Delta^{(1/6)}$ model. For the other observables in $B \to K ll$ processes we will show the results only for $\Delta^{(7/6)}$ leptoquark model.
In Fig. 3, we have shown the lepton non-universality factors  $R_K^{\mu e}$ (left panel) (\textit{i.e.} the ratio of branching ratios of 
$\bar{B} \rightarrow \bar{K} \mu^+ \mu^-$ and  $\bar{B} \rightarrow \bar{K} e^+ e^-$), $R_K^{\tau e}$ (right panel) and $R_K^{\tau \mu}$ 
(lower panel) variation with high $q^2$.    
From the figure one can see that there is significant deviations in the lepton-flavour non universality factor from their corresponding SM values
in all the above three cases.
 The flat term for the $\bar{B}_d^0 \rightarrow \bar{K}^0 \mu^+ \mu^- $ (left panel)  and 
$\bar{B}_d^0 \rightarrow \bar{K}^0 \tau^+ \tau^- $  (right panel) decay processes in the low recoil region are presented in Fig. 4 for $\Delta^{(7/6)}$.
In this case  there is practically no deviation in $B \to K \mu \mu$ whereas there is significant deviation in $B \to K \tau \tau$
process.   The integrated branching ratios, flat terms and the lepton flavour non-universality factors for the
$B \to Kll$  processes over the range $q^2 \in [14.18, 22.84]$ are
 given in Table I.  The flat term for $B \to K e^+ e^-$ process has been found to be negligibly small  ($F_H^e$ $ \sim \mathcal{O} (10^{-7})$) due to 
 tiny electron mass. In the low recoil region, the process having tau lepton in the final state has significant deviation from the SM.

The integrated branching ratio for $B^0 \to K \mu \mu$ process in the range $q^2 \in[15,22]~{\rm GeV}^2$ has been measured by the LHCb Collaboration 
\cite{lhcb1}
and is given as
\be
{\rm Br}(B^0 \to K^0 \mu \mu)=(6.7 \pm 1.1\pm 0.4) \times 10^{-8}\;.
\ee
Our predicted value in this range of $q^2$ is found to be
\bea
{\rm Br}(B^0 \to K^0 \mu \mu) &=& (8.35 \pm 0.5) \times 10^{-8},~~~~~~~~({\rm SM})\nn\\
&=&(8.34 -9.26) \times 10^{-8}\;.~~~~~~({\rm \Delta^{(7/6)}~LQ~model})\nn\\
&=&(8.34 -15.6) \times 10^{-8}\;.~~~~~~({\rm \Delta^{(1/6)}~LQ~model})
\eea
The predicted values of the branching ratios are slightly higher than the central measured value but consistent with its 1-$\sigma$ range.
 \begin{figure}[ht]
\centering
\includegraphics[scale=0.6]{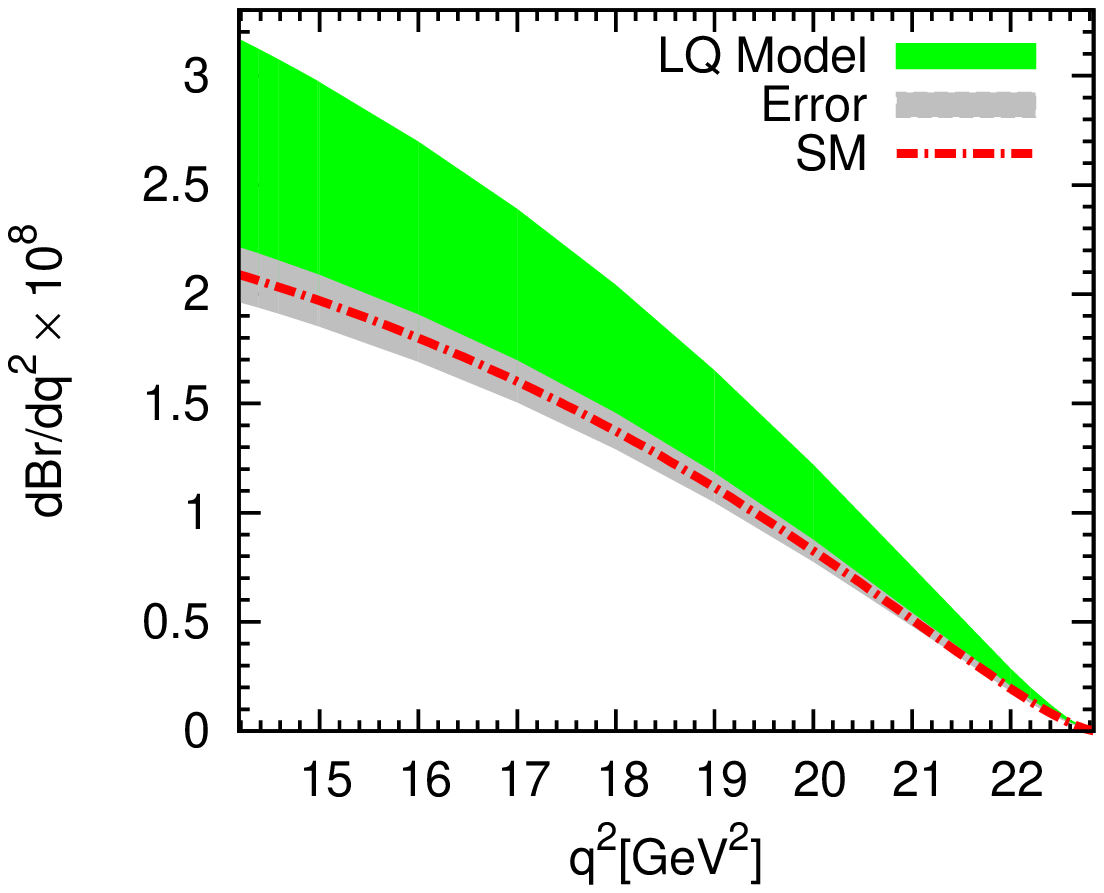}
\quad
\includegraphics[scale=0.6]{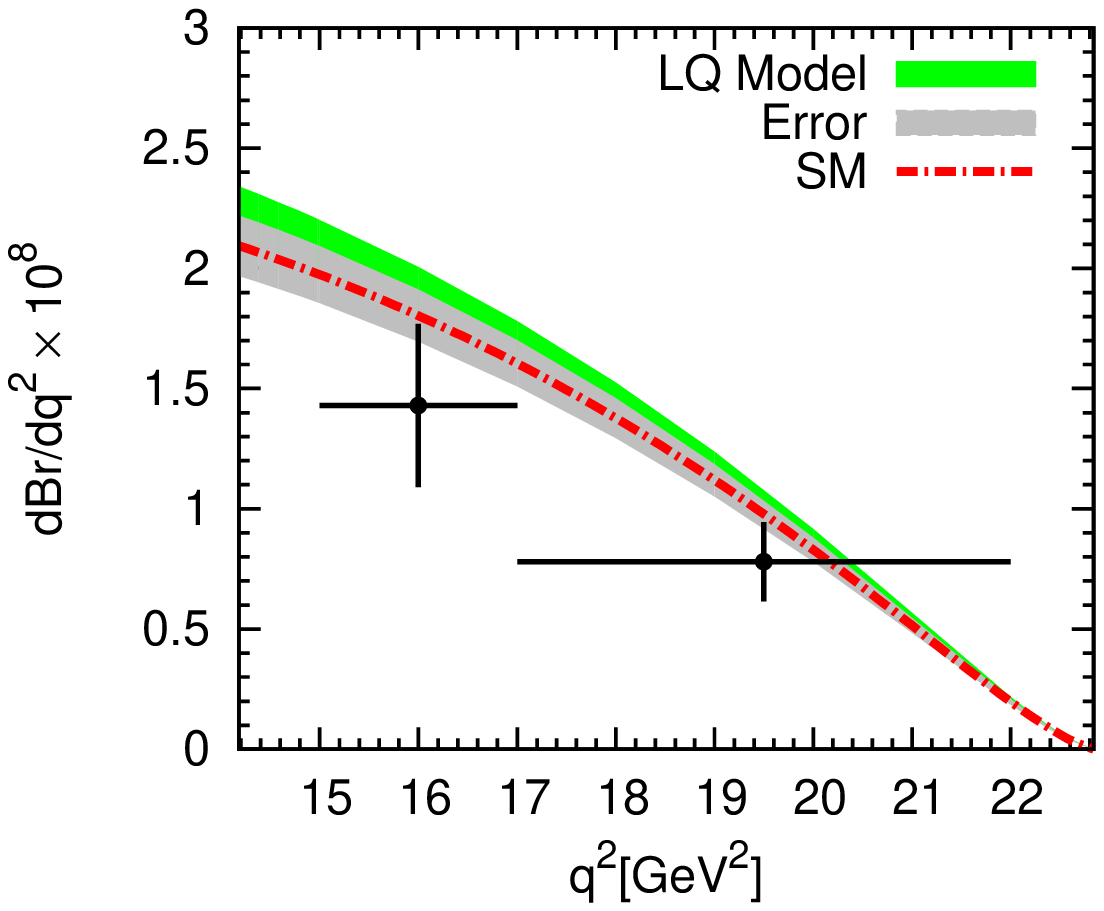}
\quad
\includegraphics[scale=0.6]{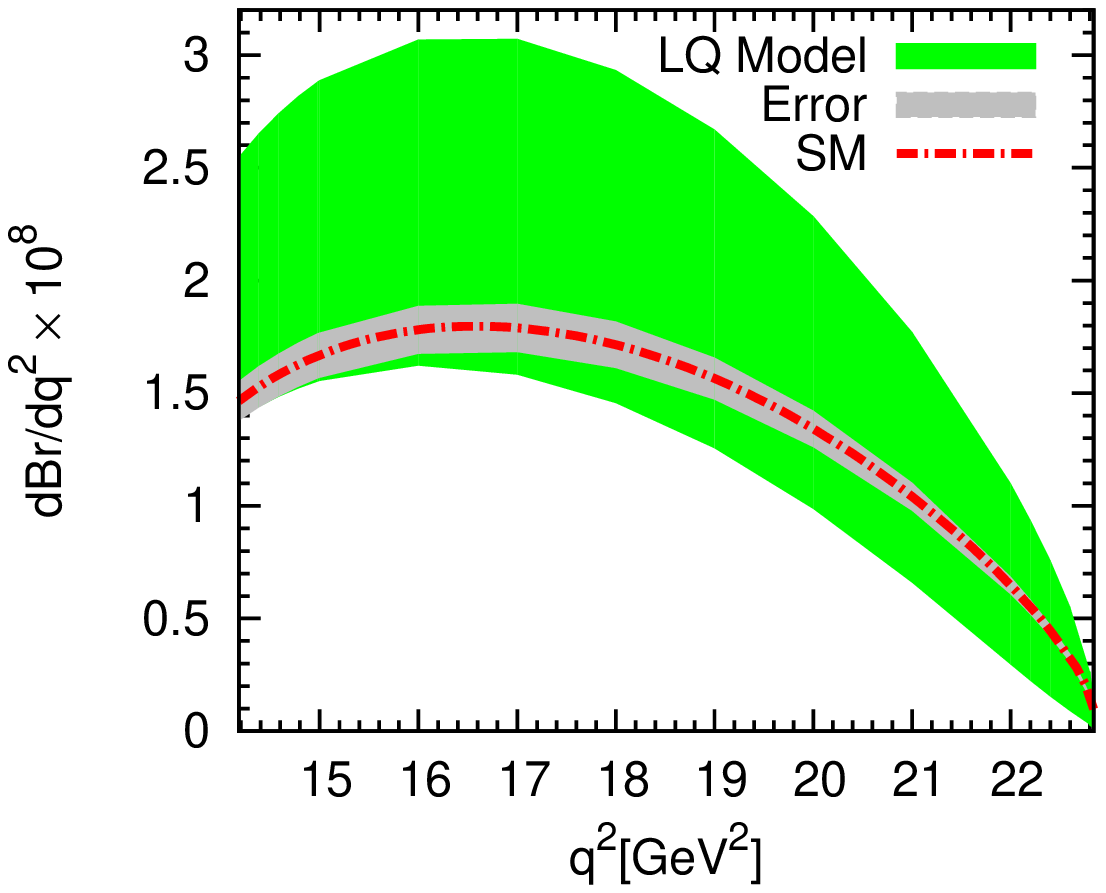}
\caption{The variation of branching ratio for $\bar{B}_d^0 \rightarrow \bar{K}^0 e^+ e^- $ (left panel), $\bar{B} \rightarrow \bar{K} \mu^+ \mu^-$ (right panel) 
and $\bar{B} \rightarrow \bar{K} \tau^+ \tau^-$ (bottom panel) with  respect to high $q^2$ for $\Delta^{(7/6)}$ LQ. 
The grey bands correspond to the uncertainties arising in the SM.
The $q^2$-averaged (bin-wise) $1-\sigma$ experimental results 
for $B \to K \mu \mu$ process are 
shown by black plots, where horizontal (vertical)
line denotes the bin width ($1-\sigma $ error).}
\end{figure}


\begin{figure}[ht]
\centering
\includegraphics[scale=0.6]{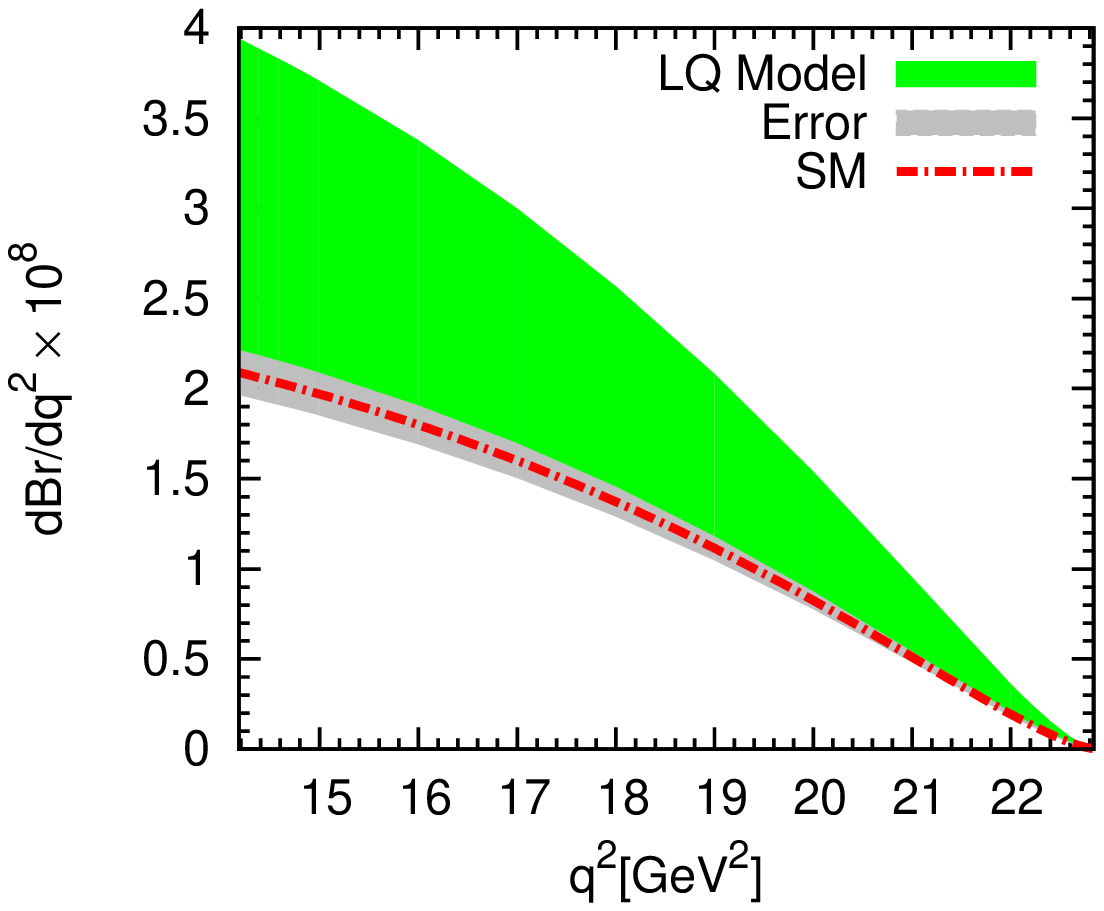}
\quad
\includegraphics[scale=0.6]{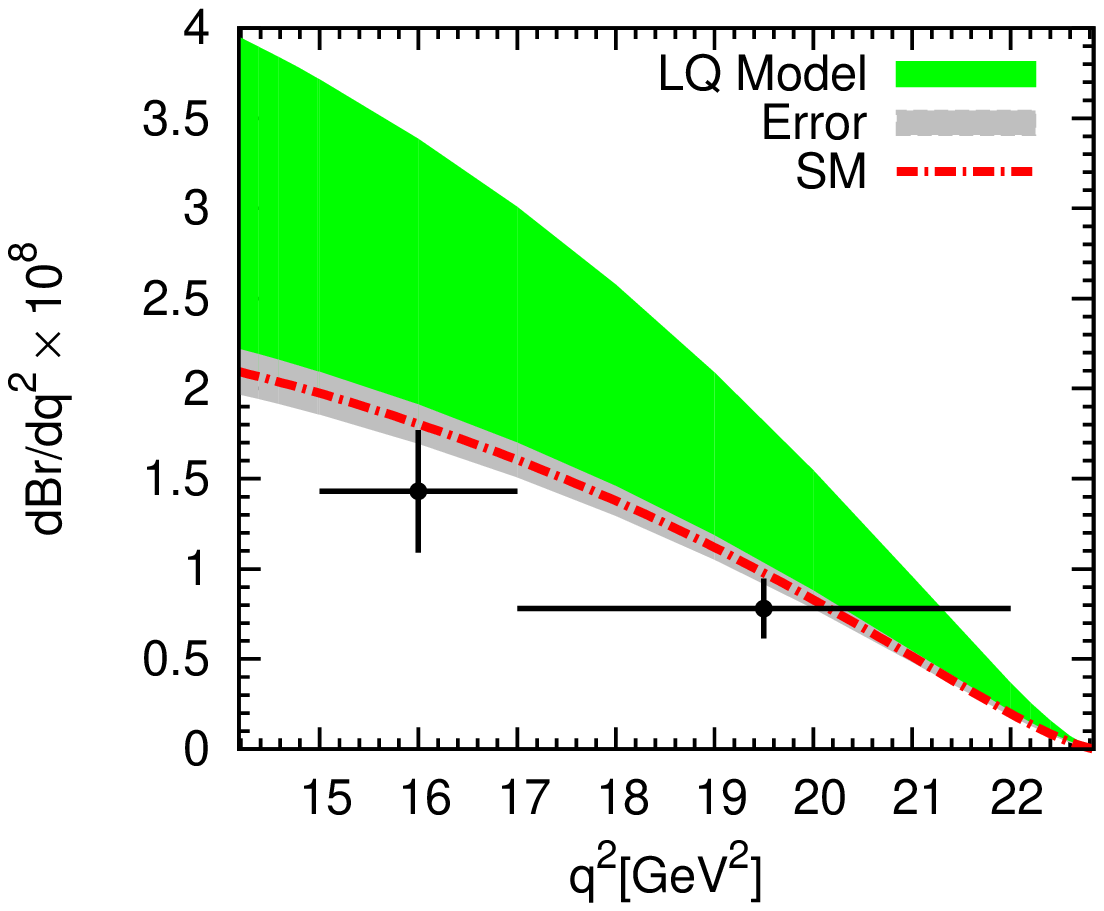}
\quad
\includegraphics[scale=0.6]{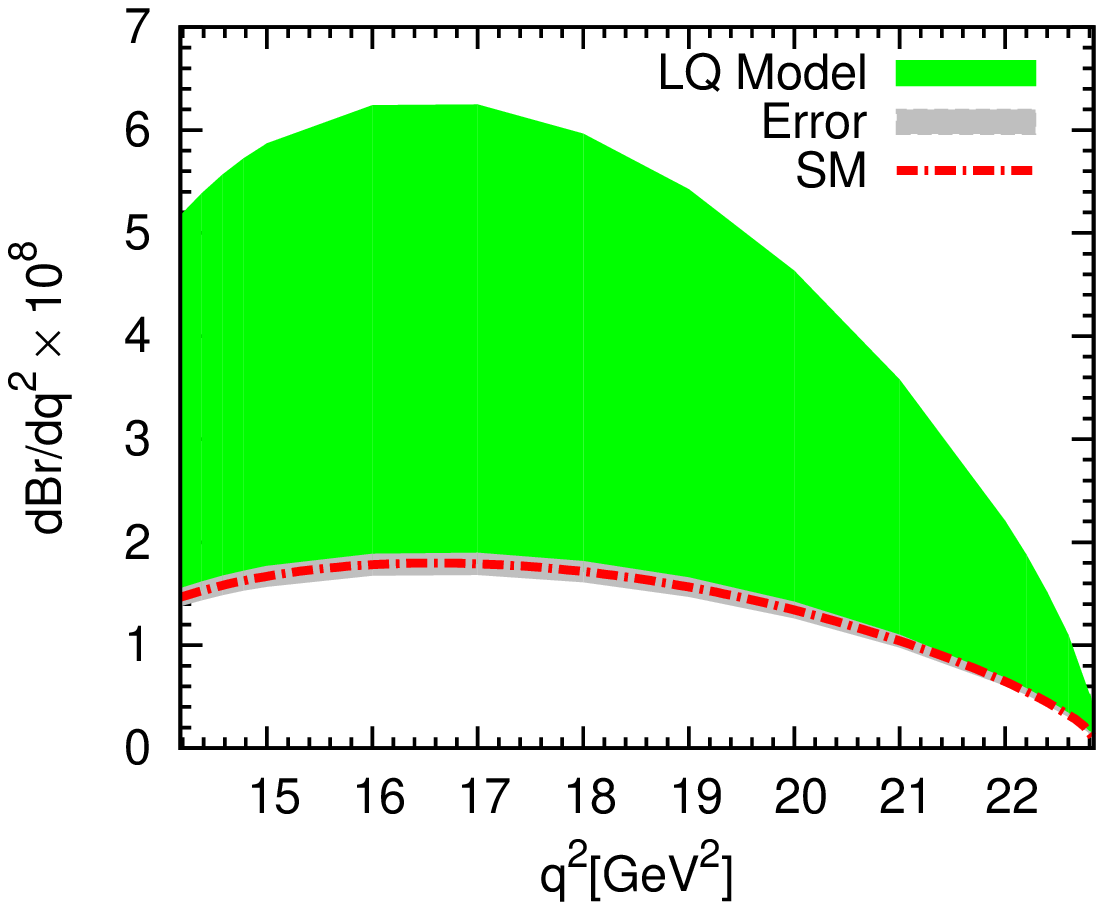}
\caption{Same as Fig-1 for  $\Delta^{(1/6)}$ LQ exhange. }
\end{figure}
 \begin{figure}[ht]
\centering
\includegraphics[scale=0.6]{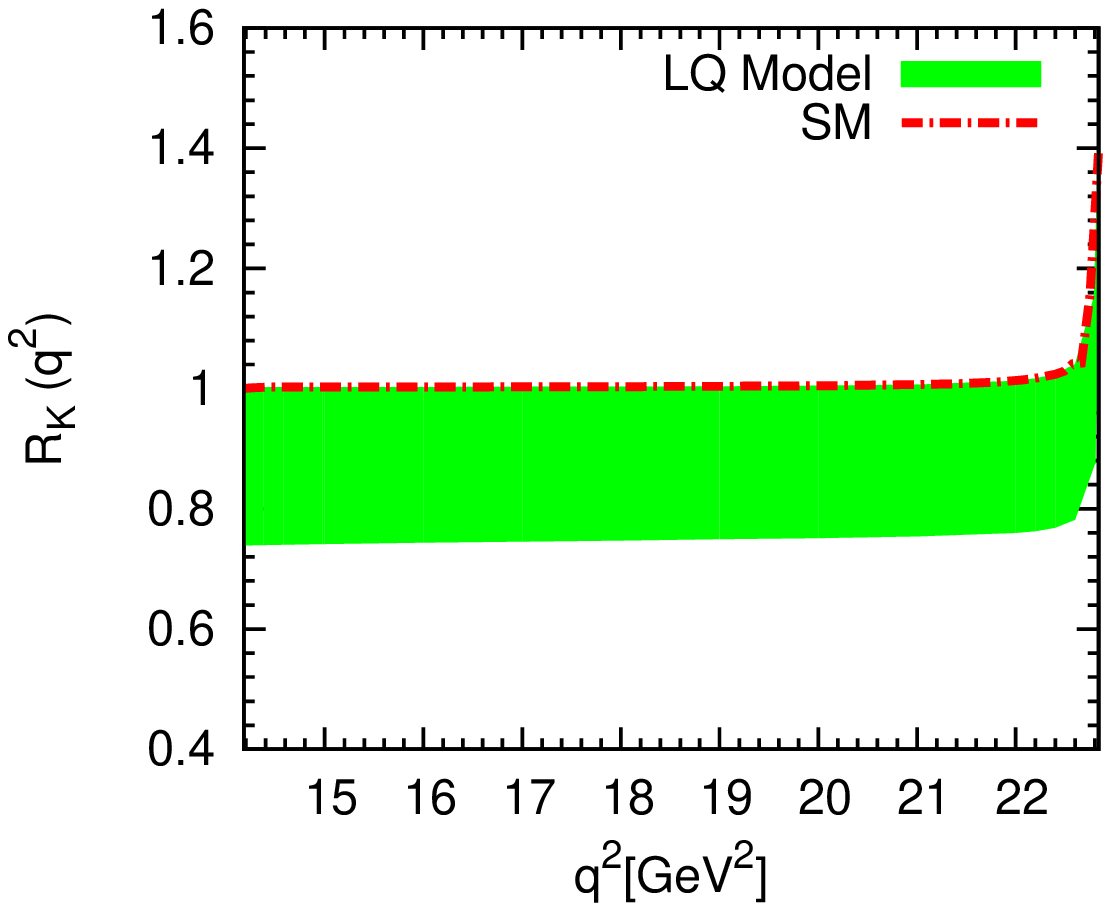}
\quad
\includegraphics[scale=0.6]{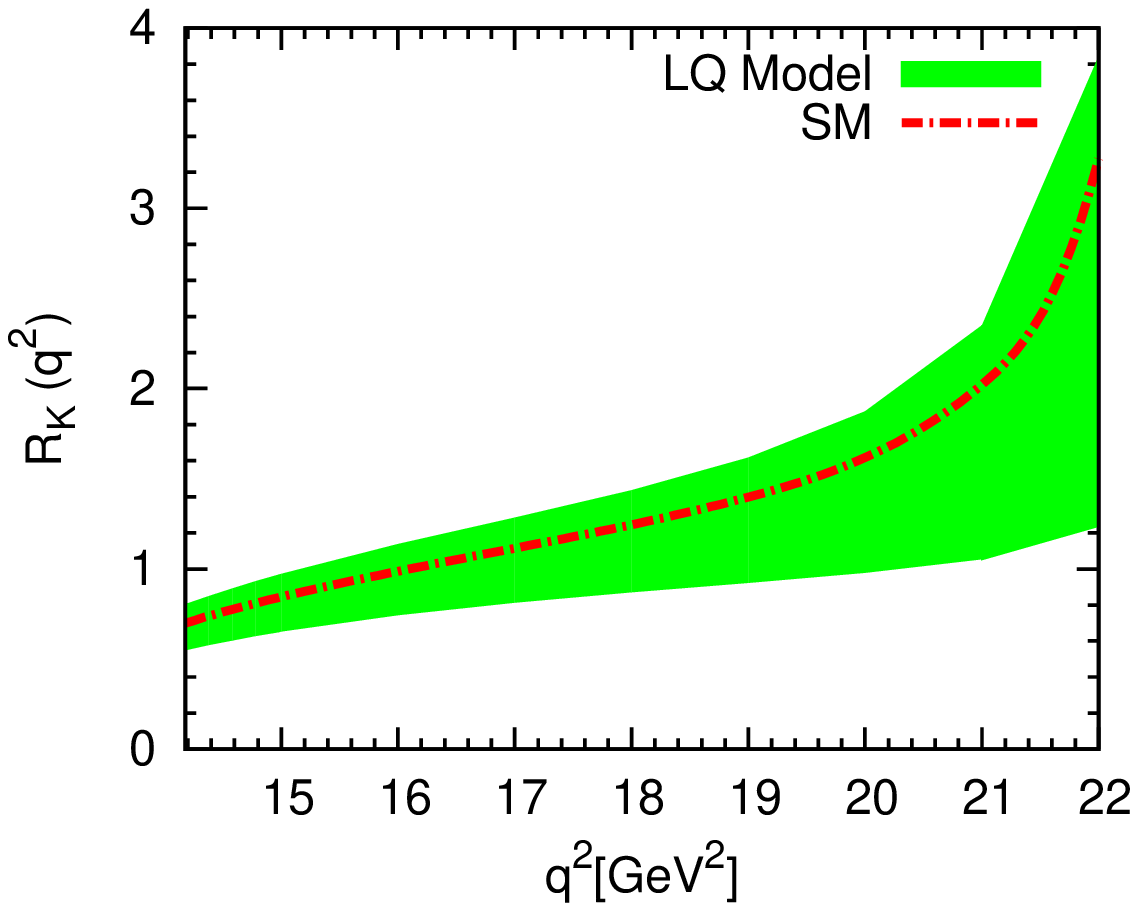}
\quad
\includegraphics[scale=0.6]{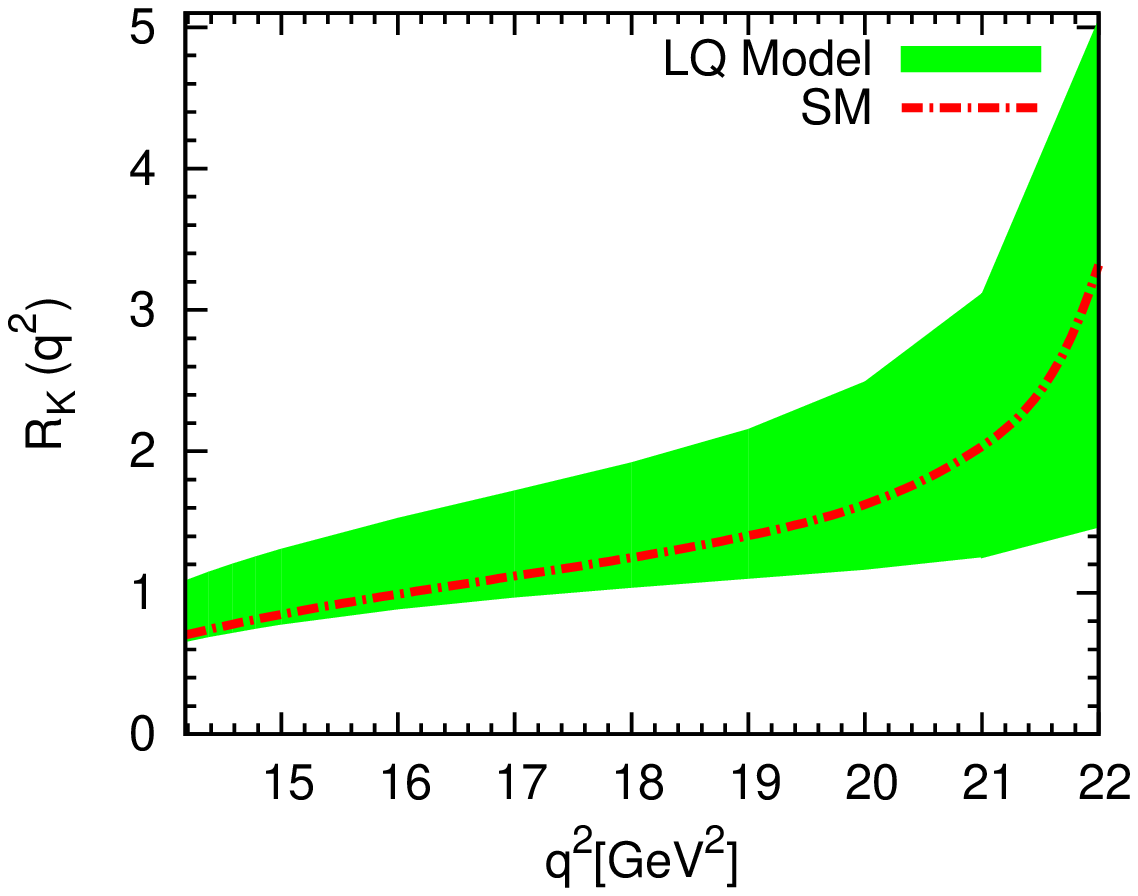}
\caption{The variation of lepton non-universality  $R_K^{\mu e}$ (left panel), $R_K^{\tau e}$ (right panel)  and $R_K^{\tau \mu}$ 
(bottom panel) in low recoil region due to $\Delta^{(7/6)}$ LQ exchange.}
\end{figure}
\begin{figure}[ht]
\centering
\includegraphics[scale=0.6]{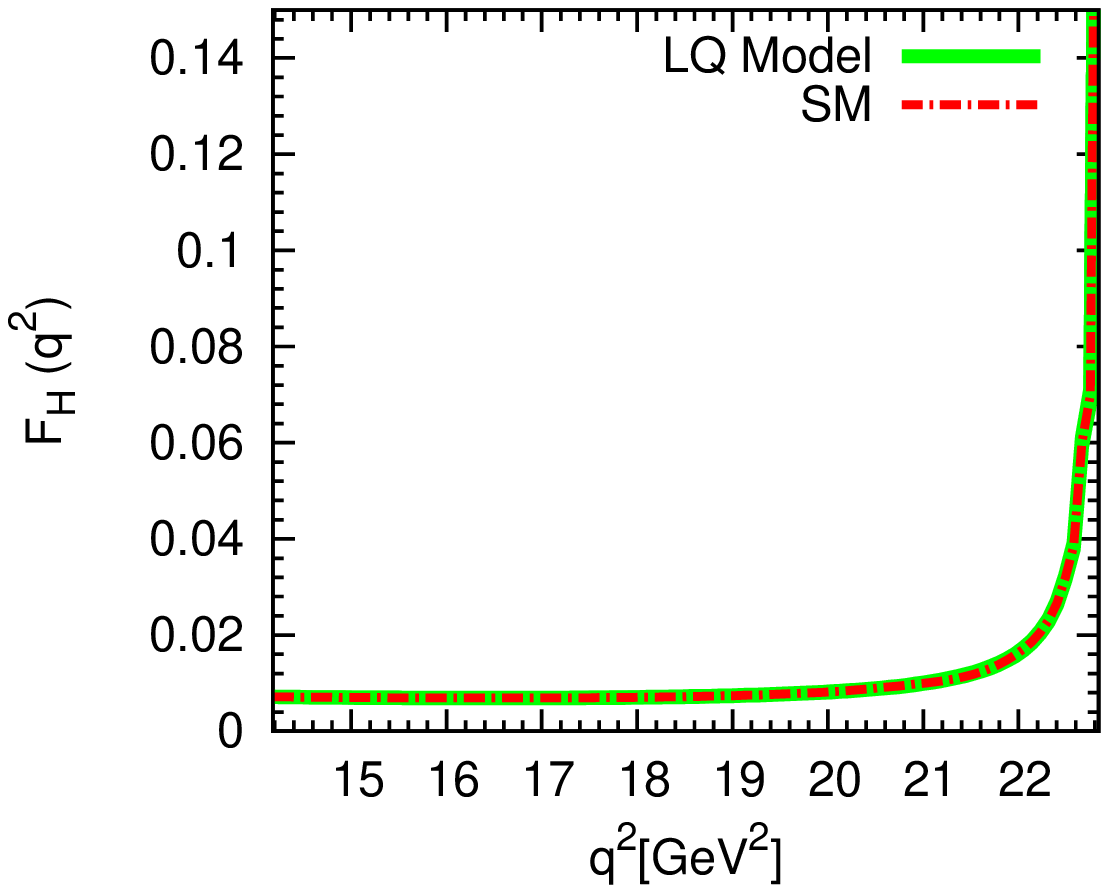}
\quad
\includegraphics[scale=0.6]{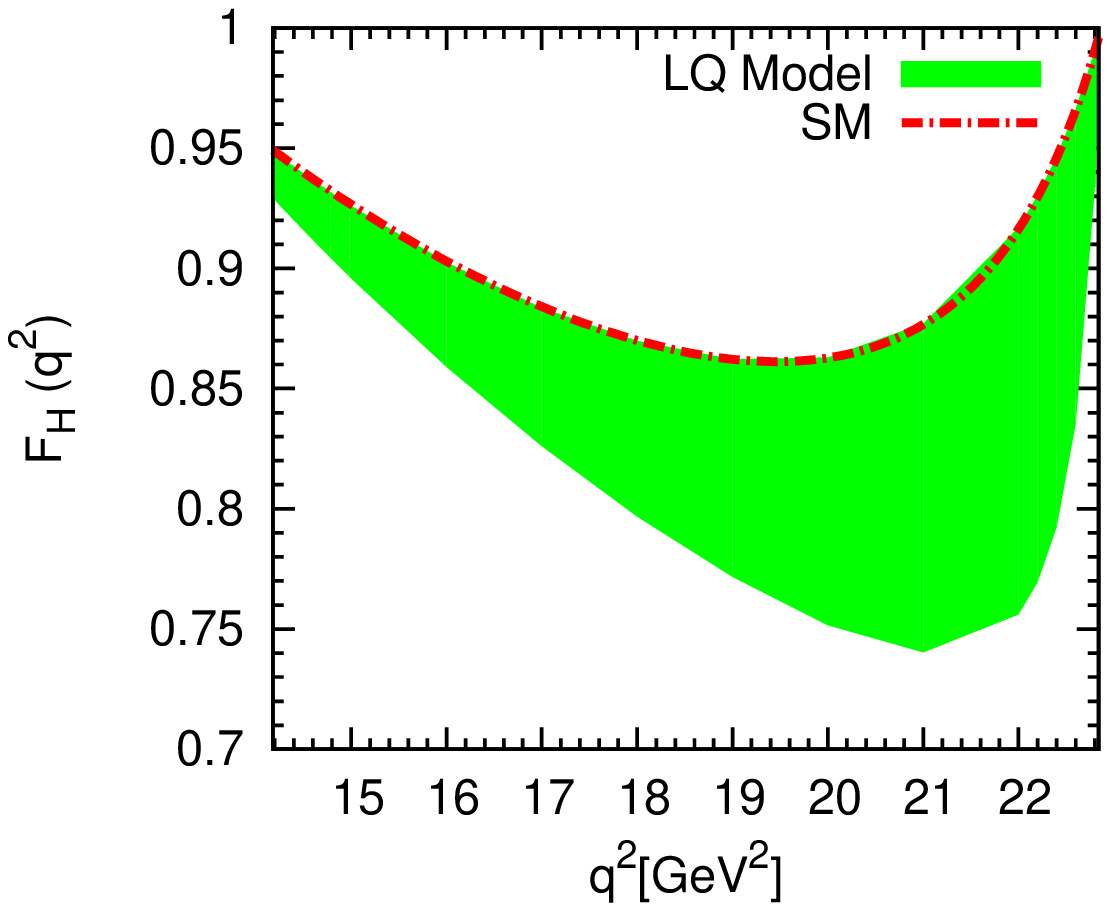}
\caption{The variation of flat term for $\bar{B}_d^0 \rightarrow \bar{K}^0 \mu^+\mu^-$ (left panel) and $\bar{B}_d^0 \rightarrow \bar{K}^0 \tau^+\tau^-$ 
(right panel) with high $q^2$ for $\Delta^{(7/6)}$ LQ.}
\end{figure}

\begin{table}[ht]
\begin{center}
\begin{tabular}{|c c c c |}
\hline
 Oservables ~&~ SM  predictions ~&~ Values in $\Delta^{(7/6)}$ LQ model ~ & ~ Values in  $\Delta^{(1/6)}$ LQ model \\
   \hline
 
Br($B_d^0 \to K^0 e^+ e^-$) & $(1.005\pm 0.06)$  & $(1.004 - 1.5)$  &  $(1.005 - 1.88)$   \\

Br($B_d^0 \to K^0 \mu^+ \mu^-$) & $(1.01\pm 0.06)$  & $(1.01 - 1.12)$ & $(1.008 - 1.89)$   \\
Br($B_d^0 \to K^0 \tau^+ \tau^-$) & $(1.21\pm 0.73)$  & $(0.99 - 2.07)$  & $(1.2 - 4.2)$ \\
$\langle R_K^{\mu e} \rangle$ & 1.0035 & $0.75-1.00$ & 1.0035  \\
$\langle R_K^{\tau e} \rangle$ & 1.21 & $0.98 - 1.85$ & $(1.2 - 2.3)$ \\
 $\langle R_K^{\tau \mu} \rangle$ & 1.198 & $0.98 - 1.85$ & $(1.2 - 2.2)$ \\
$\langle F_H^e \rangle$ & $1.75 \times 10^{-7}$ & $(1.74 - 1.75) \times 10^{-7}$ &  $ ( 1.73 - 1.75)\times 10^{-7}$ \\
$\langle F_H^\mu \rangle$ & $7.5 \times 10^{-3}$  & $(7.4 - 7.55)\times 10^{-3}$  & $(7.4 - 7.5)\times 10^{-3}$ \\
$\langle F_H^\tau \rangle$ & 0.89 & 0.8 - 1.38 & 0.88-0.89 \\

 \hline
\end{tabular}
\end{center}
\caption{The predicted values for the integrated branching ratios (in units of $10^{-7}$), flat terms and lepton non-universality factors in 
the range $q^2 \in [14.18,22.84]~ {\rm GeV}^2$ for the $B \to K l^+ l^-$ process, $l = e, \mu, \tau$. }

\end{table}
 
\section{$B \rightarrow K \nu \bar{\nu}$ process}
The $B \rightarrow K \nu \bar{\nu}$ process is mediated by the quark level transition $b \to s \nu \bar{\nu}$ and
the effective Hamiltonian describing such transition  is given as  \cite{buras2} 
\begin{equation}
\mathcal{H}_{eff} = \frac{-4G_f}{\sqrt{2}} V_{tb}V_{ts}^*\left(C^\nu_L \mathcal{O}^\nu_L +C^\nu_R \mathcal{O}^\nu_R \right) + h.c., \label{nu1}
\end{equation}
where 
\begin{equation}
\mathcal{O}^\nu_L = \frac{e^2}{16 \pi^2} \left(\bar{s}\gamma_\mu P_L b \right) \left(\bar{\nu} \gamma^\mu \left(1-\gamma_5\right)\nu\right),
 \hspace{1cm} \mathcal{O}^\nu_R = \frac{e^2}{16 \pi^2} \left(\bar{s}\gamma_\mu P_R b \right) \left(\bar{\nu} \gamma^\mu \left(1-\gamma_5\right)\nu\right),
\label{nu-ham}
\end{equation}
are the dimension-six operators and $C^\nu_{L,R}$ are their corresponding Wilson coefficients. The   coefficient $C^\nu_R$ 
has negligible value within the standard model while $C^\nu_L$ can be calculated by using the loop function and is 
given by 
\bea
 C^\nu_L = -X(x_t)/\sin^2\theta_w\;. \label{cl}
\eea
The necessary loop functions are presented in Appendix B.
The decay distribution with respect to the di-neutrino invariant mass can be expressed as \cite{fazio}
\begin{equation}
\frac{d\Gamma}{ds_B} = \frac{G_f^2\alpha^2}{256\pi^5} |V_{ts}^* V_{tb}|^2 m_B^5 \lambda^{3/2}(s_B,\tilde{m}_K^2,1) |f_+^K(s_B)|^2 |C_L^\nu + C_R^\nu|^2.
\label{nu} 
\end{equation}
where $\tilde{m}_i = m_i/m_B$ and $s_B = s/m_B^2$.  The decay rate has been multiplied with an extra factor 3 due to the sum over all neutrino flavours.
It should be noted that  in Eq. (\ref{nu}) $C_R^\nu$ is the new Wilson coefficient arises due to the exchange  of the leptoquark $\Delta^{(1/6)}$. 
In order to find out its value, we consider  the new contribution to 
the effective Hamiltonian due to the exchange of such leptoquark which is given as
\begin{equation}
 {\cal H}_{LQ}=\frac{(g_L)_{s \nu} (g_L)_{b \nu}^*}{4 M_{\Delta^{(1/6)}}^2} (\bar s \gamma^\mu P_R b)(\bar{\nu} \gamma_\mu (1-\gamma_5) \nu)\;.\label{nu2}
\end{equation}
Comparing Eqs. (\ref{nu1}) and (\ref{nu2}), one can obtain the new Wilson coefficient as
\begin{equation}
 C_R^\nu|_{LQ}= -\frac{\pi}{2 \sqrt 2 G_F \alpha V_{tb} V_{ts}^*}\frac{(g_L)_{s \nu} (g_L)_{b \nu}^*}{M_{\Delta^{(1/6)}}^2}\;.
\end{equation}
For numerical estimation, we use the $B \to K$ form factor $f_+^K$ evaluated in the light cone sum rule (LCSR)  approach \cite{ball3} as
\begin{equation}
f_+^K (q^2) = \frac{r_1}{1-q^2/m_1^2} +  \frac{r_2}{(1-q^2/m_1^2)^2} \;,
\end{equation} 
which is valid in the full physical region. Furthermore, in contrast to  $B \to K l^+ l^-$  process, which has dominant charmonium resonance background 
from $B \to K (J/\psi) \to K l^+ l^-$, there are no such analogous long-distance QCD contributions in this case as there are no 
intermediate states which can decay into two neutrinos. For the $b\to s \nu \bar \nu $ LQ couplings we use the values as we used for $b \to s \mu \mu$ as these 
two processes are related by $SU(2)_L$ symmetry. The variation of branching ratio with respect to $s_B$ in the full physical regime 
$0 \leqslant s_B \leqslant (1-\tilde{m}_K)^2$ is shown in Fig. 5 and the predicted branching ratio is given in Table II, which is well below the present
upper limit ${\rm Br}(B_d^0 \to K \nu \bar \nu) < 4.9 \times 10^{-5}$ \cite{pdg}.
\begin{figure}[ht]
\centering
\includegraphics[scale=0.6]{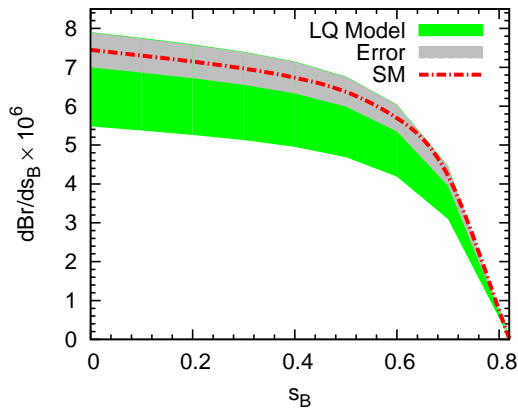}
\caption{The variation of branching ratio of $B \rightarrow K \nu \bar{\nu}$ with respect to the normalized invariant masses squared $s_B$
in the SM and $\Delta^{(1/6)}$ LQ model.
The grey band corresponds to the uncertainties arising in the SM.}
\end{figure}
\section{$B \to K^* \nu \bar{\nu}$ process}
The study of $B \to K^* \nu \bar \nu$ is also quite important as this process is related to $B \to K^* \mu \mu$ process by $SU(2)_L$ and therefore,
the recent LHCb anomalies in $B \to K^* \mu \mu$ would in principle also show up in $B \to K^* \nu \nu$.
The experimental information about this  exclusive decay process  can be described by the double differential decay distribution. 
In order to compute the decay rate, we must have the  idea about the matrix element of the effective Hamiltonian 
(\ref{nu1}) between the initial $B$ meson and the final 
particles. Due to the non-detection of the two neutrinos, experimentally we can't distinguish between the transverse polarization, so the decay rate will 
be the addition of both longitudinal and transverse polarizations. 
The double differential decay rate with respect to $s_B$ and $\cos\theta$ is given by \cite{buras2, kim}
\begin{equation}
\frac{d^2 \Gamma}{ds_B d \cos \theta} = \frac{3}{4} \frac{d \Gamma_T}{ds_B } \sin^2 \theta + \frac{3}{2} \frac{d \Gamma_L}{ds_B } \cos^2 \theta,
\end{equation}
     where the longitudinal and transverse decay rate are 
\begin{equation}
\frac{d \Gamma_L}{ds_B } = 3 m_B^2 |A_0|^2,  \hspace{2cm} 
\frac{d \Gamma_T}{ds_B } = 3 m_B^2 (|A_\perp|^2  + |A_\parallel|^2).
\end{equation}
The transversality amplitudes $A_{\perp, \parallel, 0}$ in terms of the  form factors and Wilson coefficients are listed in
Appendix C.

The fractions of $K^*$ longitudinal and transverse  polarizations are given as 
\begin{equation}
F_{L,T} = \frac{d \Gamma_{L,T}/ds_B}{d \Gamma /ds_B}, 
\end{equation}
and the $K^*$ polarization factor is
\begin{equation}
\alpha_{K^*} = \frac{2 F_L}{F_T} -1.
\end{equation}
The transverse asymmetry parameters are given as \cite{egede, simula}
\begin{equation}
A_T^{(1)} = \frac{-2 Re(A_\perp A_\parallel^*) }{ |A_\perp|^2 +|A_\parallel|^2 }\;,~~~~~~~~
A_T^{(2)} =  \frac{|A_\perp|^2  - |A_\parallel|^2 }{|A_\perp|^2 +|A_\parallel|^2 }.
\end{equation}
However, one can't extract $A_T^{(1)}$  from the full angular distribution of $B \rightarrow K^* \nu \bar{\nu}$, as it is not invariant under the symmetry of 
the distribution function and requires measurement of the neutrino polarization. So it can't be measured experimentally at $B$ factories or in LHCb.  
The transverse asymmetry $A_T^{(2)}$ is theoretically clean and could be measurable in Belle-II. 

For numerical evaluation we use the $q^2$ dependence of the  $B \to K^*$ form factors  $V(q^2), ~ A_1(q_2),~ A_2 (q^2)$  from \cite{ball1, ball2}.
The variation of the  branching ratio of $B \rightarrow K^* \nu \bar{\nu}$ with respect to the 
neutrino invariant mass, $s_B$ is shown in Fig. 6. Fig. 7 contains the longitudinal and transverse polarizations of $K^*$ verses $s_B$. 
The polarization factor and the transverse asymmetry variation with respect to $s_B$ in the full region are shown in Fig. 8. 
Although there is certain deviation found between the SM and LQ model predictions for the branching fraction, but no such noticeable
deviations found between the SM and LQ predictions for the  
 longitudinal/transverse
polarizations, transverse asymmetry parameters $A_T^{(2)}$. 
The integrated values of branching ratio over the range  $s_B \in [0, 0.69]$ are presented in Table II, which are well below the
the present upper limit ${\rm Br}(B_d^0 \to K^* \nu \bar \nu) < 5.5 \times 10^{-5}$ \cite{pdg}.

\begin{figure}[ht]
\centering
\includegraphics[scale=0.6]{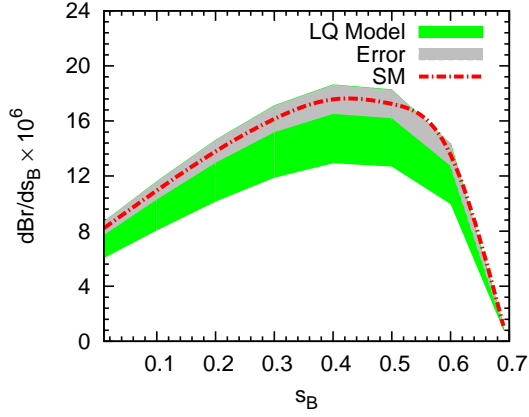}
\caption{The variation of branching ratio of $B \rightarrow K^* \nu \bar{\nu}$ with respect to the $s_B$ in the SM and $\Delta^{(1/6)}$ LQ model.
The grey band corresponds to the uncertainties arising in the SM.}
\end{figure}
\begin{figure}[ht]
\centering
\includegraphics[scale=0.6]{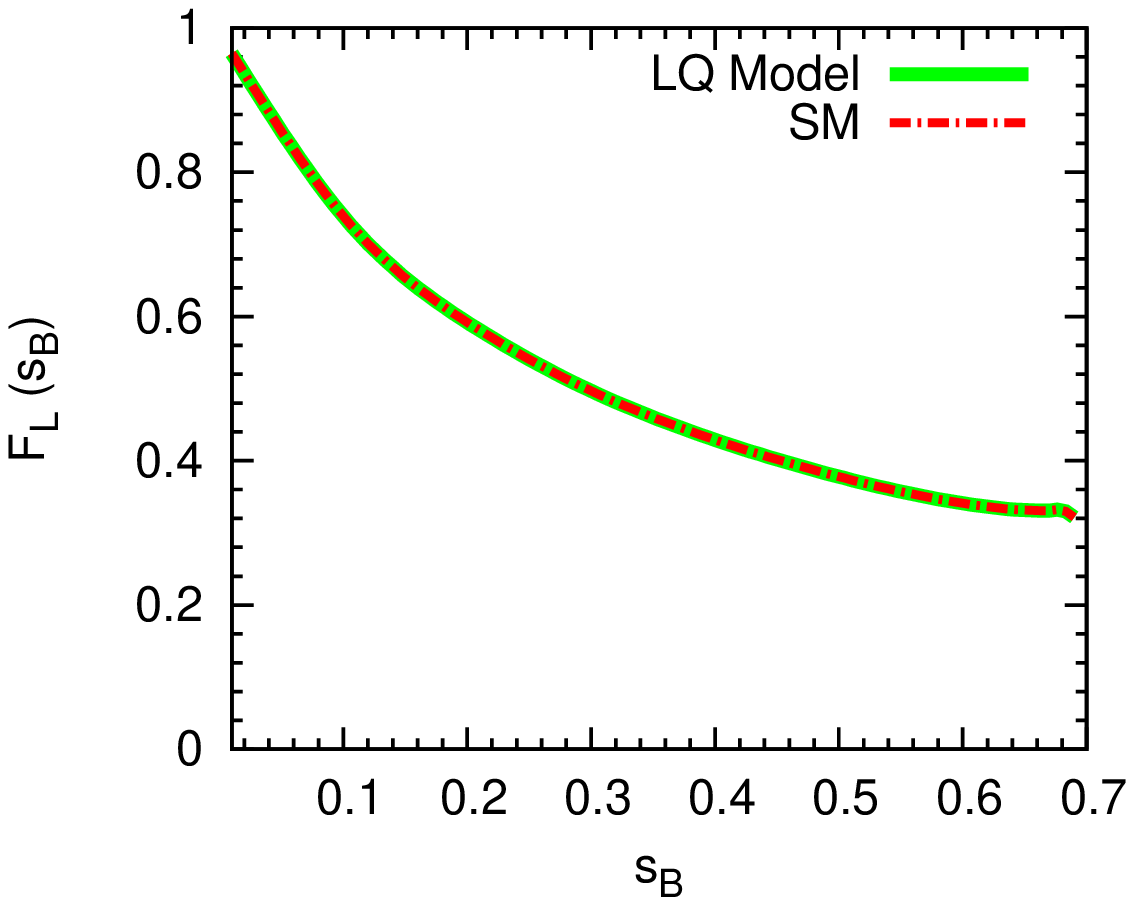}
\quad
\includegraphics[scale=0.6]{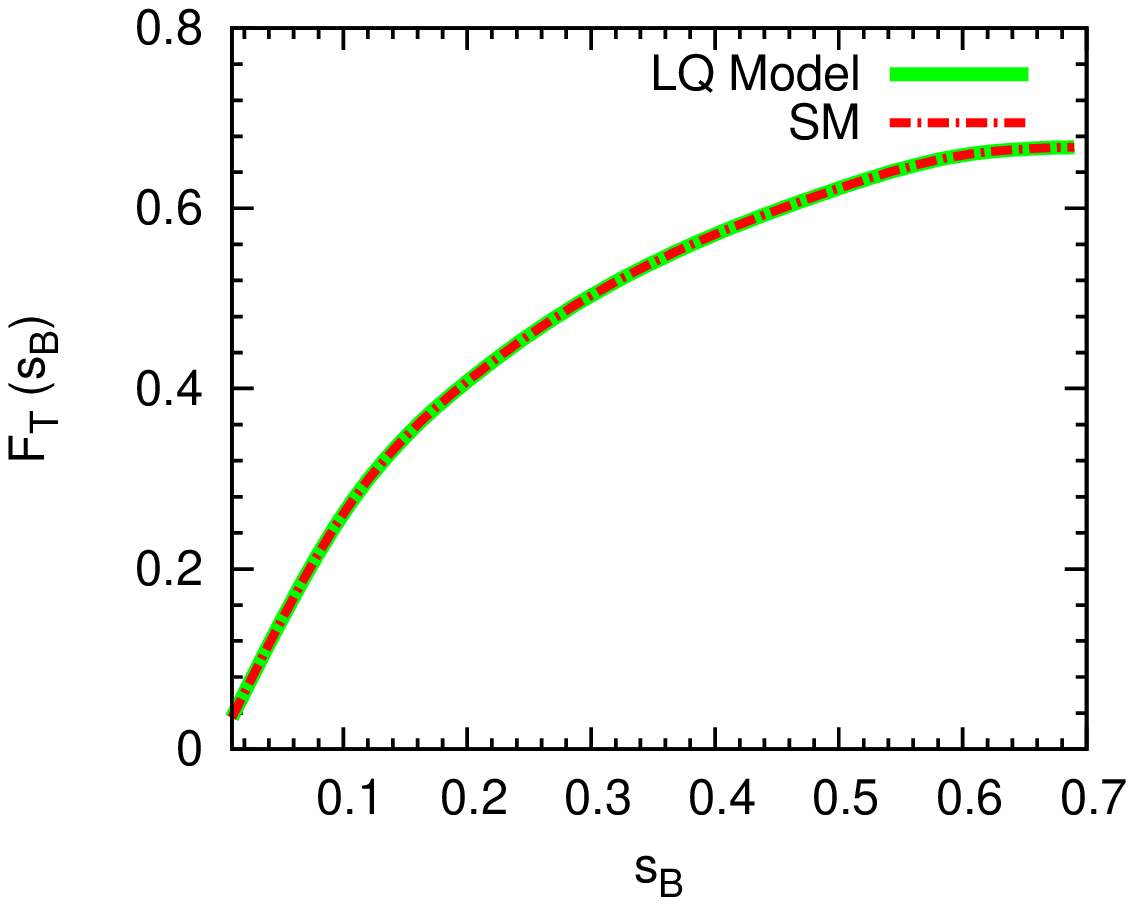}
\caption{The variation of longitudinal (left panel) and transverse (right panel) polarization of $K^*$ with $s_B$.}
\end{figure}
\begin{figure}[ht]
\centering
\includegraphics[scale=0.6]{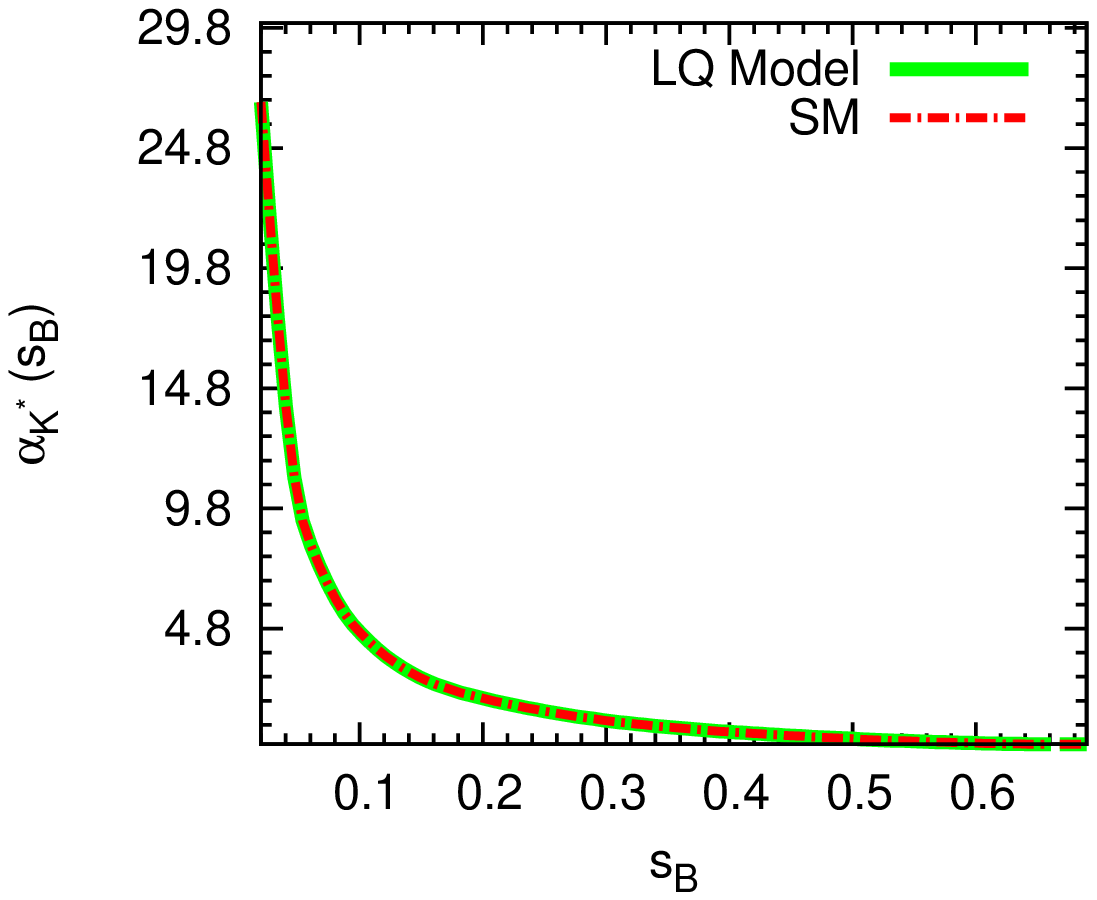}
\quad
\includegraphics[scale=0.6]{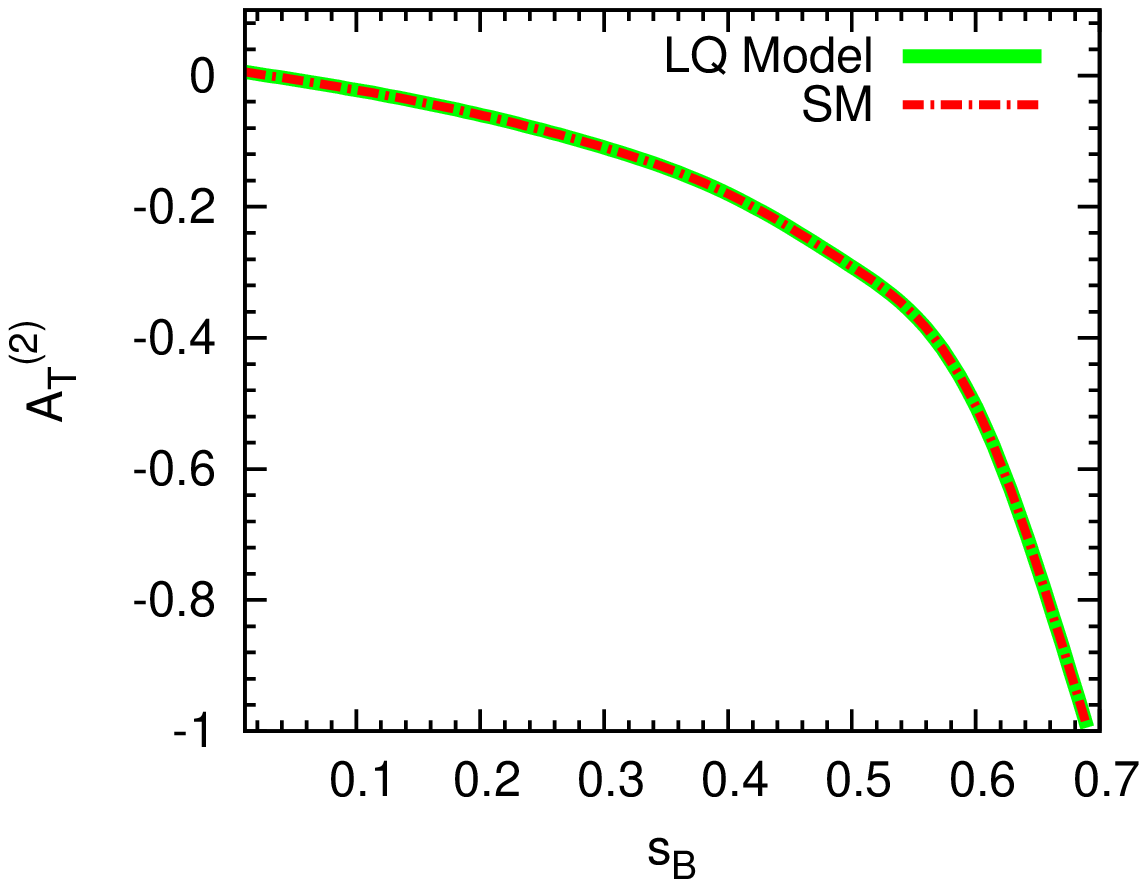}
\caption{The variation of $K^*$ polarization factor (left panel) and the transverse asymmetry (right panel)  with  respect to $s_B$.}
\end{figure}
\section{$B \to X_s \nu \bar{\nu}$}

The inclusive decay $B \to X_s \nu \bar{\nu}$ is dominated by the $Z$-exchange and can be searched through the large missing energy 
associated with the two neutrinos. This  decay mode is theoretically very clean, since both the perturbative $\alpha_s$ 
and the non-perturbative corrections are small. So these decays do not suffer  from the form factor uncertainties and 
thus, are very sensitive to the  search for new physics beyond  the SM. 
The decay distribution with respect to $s_b=s/m_b^2$ can be written as
\begin{eqnarray}
\frac {d\Gamma}{d s_b} &= & m_b^5 \frac{\alpha^2 G_f^2}{128 \pi^5} |V_{ts}^* V_{tb}|^2 \kappa(0) (|C_L^\nu|^2 + |C_R^\nu|^2)  
\lambda^{1/2} (1,\tilde{m}_s^2, s_b)\nn \\ &  \times&
\left[3s_b (1+ \tilde{m}_s^2 - s_b-4\tilde{m}_s \frac{Re(C_L^\nu C_R^{\nu *})}{|C_L^\nu|^2 + |C_R^\nu|^2}) + \lambda(1,\tilde{m}_s^2, s_b)  \right]
\end{eqnarray}
where $\tilde{m}_i = m_i/m_b$ and $\kappa(0) = 0.83$ is the QCD correction to the $b \to s \nu \bar{\nu}$ matrix element \cite{nardi}. 
The full kinematically accessible physical region is $0 \leq s_b \leq (1-\tilde{m}_s)^2$. In Fig. 9, we have shown the variation of the branching ratio
 with respect to $s_b$ and the integrated branching ratio values over the range  $s_b \in [0, 0.96]$ both in the SM and in the leptoquark model 
are presented in Table II.

We define the ratio of branching ratios as \cite{fazio},
\begin{equation}
R_K = \frac{{\rm Br} (B \to K \nu \bar{\nu})}{{\rm Br} (B \to X_s \nu \bar{\nu})},
\end{equation}
 and 
\begin{equation}
R_{K^*}= \frac{{\rm Br} (B \to K^* \nu \bar{\nu})}{{\rm Br} (B \to X_s \nu \bar{\nu})}
\end{equation} 
and the variation of $R_K$ and $R_{K^*}$ with respect to $s_B$ in the full kinematically allowed region is shown in Fig. 10. In this case also
no deviation found between the SM and leptoquark predictions.

\begin{figure}[ht]
\centering
\includegraphics[scale=0.6]{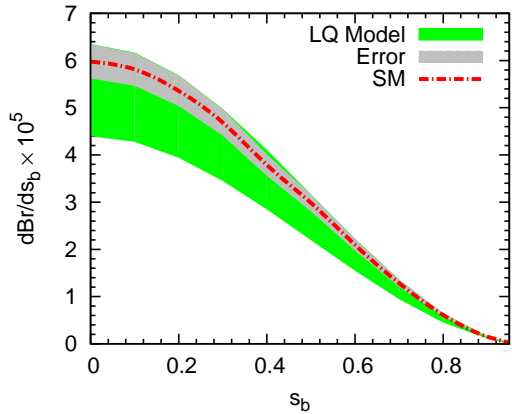}
\caption{The variation of branching ratio of $B \rightarrow X_s \nu \bar{\nu}$ with respect to the $s_b$.}
\end{figure}
\begin{figure}[ht]
\centering
\includegraphics[scale=0.6]{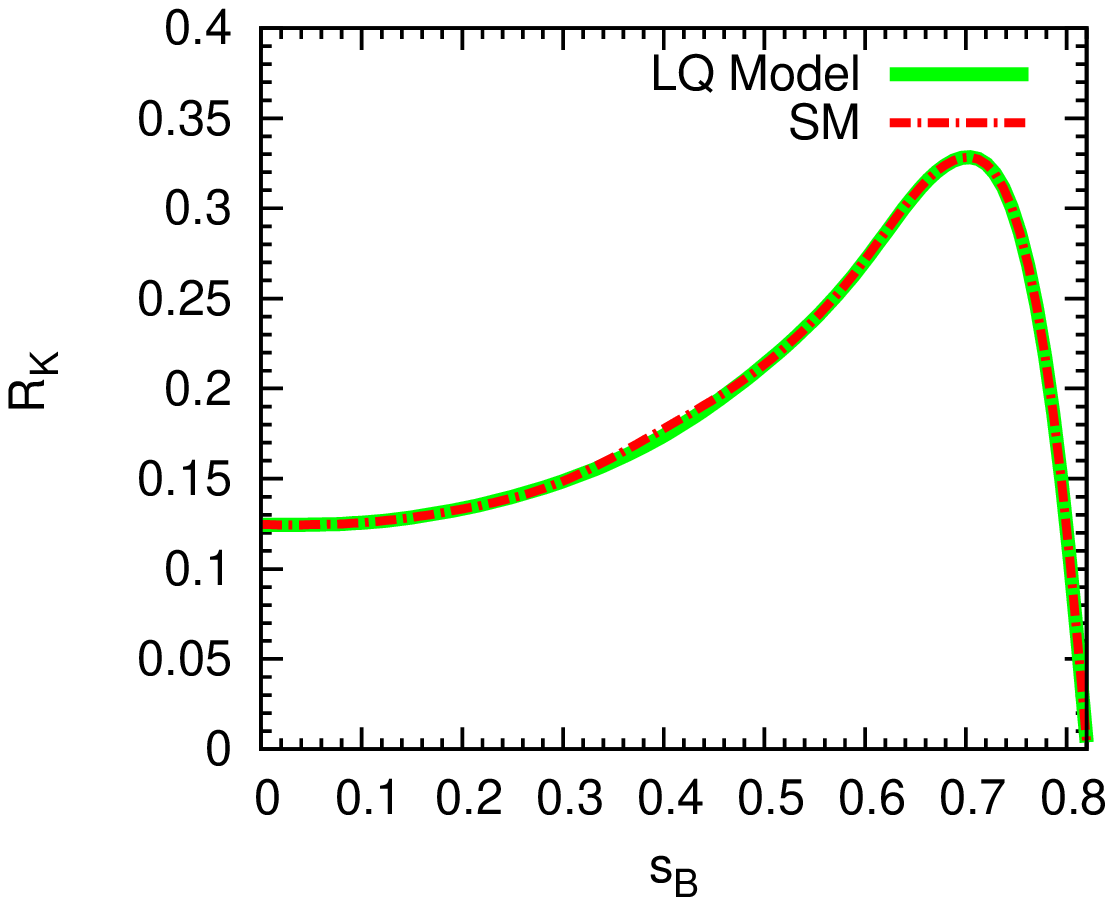}
\quad
\includegraphics[scale=0.6]{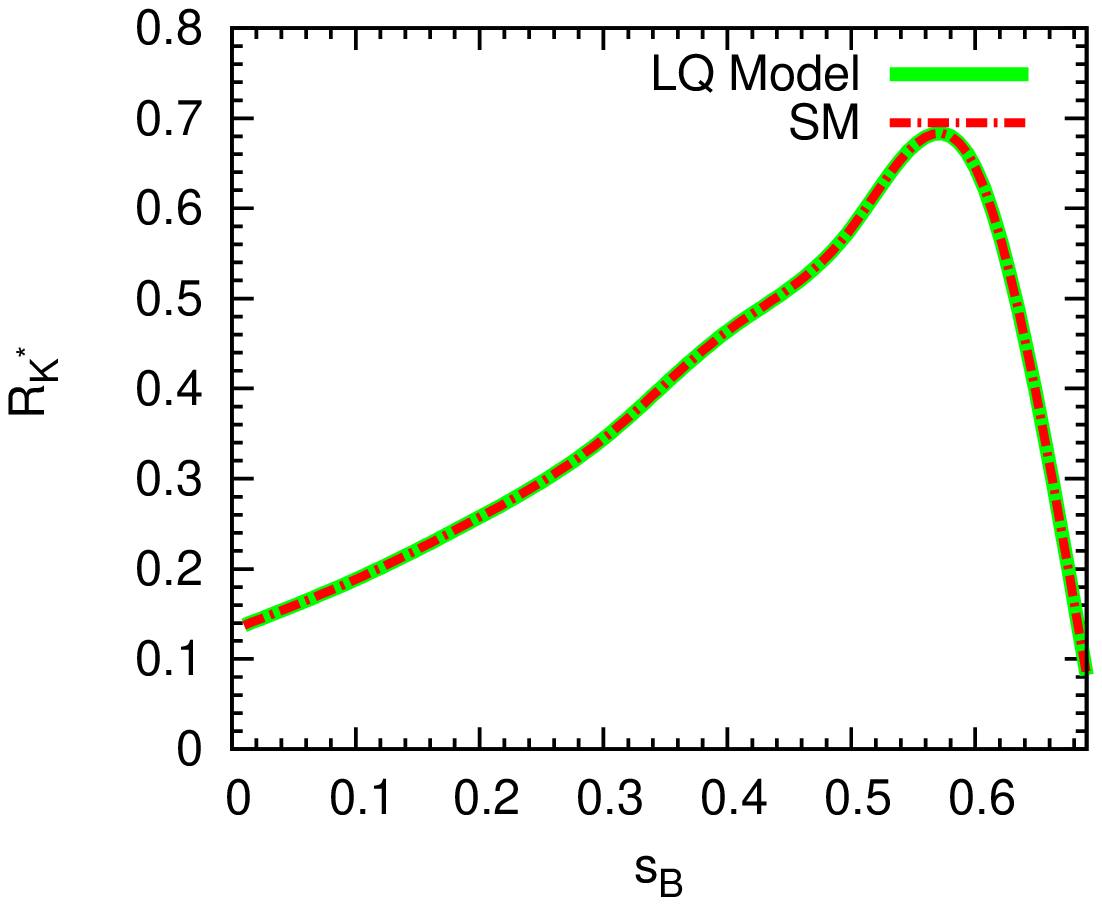}
\caption{The variation of $R_K$ (left panel) and $R_{K^*}$ (right panel)  with  respect to $s_B$.}
\end{figure}


\begin{table}[ht]

\begin{center}
\begin{tabular}{c c c}
\hline

 Observables ~&~ SM prediction~ &~ Values in $\Delta^{(1/6)}$ LQ model\\
 \hline
 \hline
Br($B_d^0 \to K^0 \nu \bar{\nu}$) & ~~$(4.9\pm 0.29)\times 10^{-6}$ ~~& $(3.6- 5.2)\times 10^{-6}$\\
Br($B_d^0 \to K^* \nu \bar{\nu}$) & $(9.54 \pm 0.57)\times 10^{-6}$ & $(7.02 - 10.13) \times 10^{-6}$\\
Br($B \to X_s \nu \bar{\nu}$) & $(2.98\pm0.18) \times 10^{-5}$ & $ (2.2 - 3.17)\times 10^{-5}$\\
$R_K$ & 0.164 & $(0.163-0.164)$\\
$R_{K^*}$ & 0.32 & 0.32\\ 
 \hline
\end{tabular}
\end{center}
\caption{The predicted branching ratios for $B \to {(K, K^*, X_s)} \nu \bar{\nu}$ processes and $R_{K,K^*}$ for $B \to X_s \nu \bar \nu$ 
in  their  respective full  physical ranges.  }
\end{table}

\section{conclusion}
In this paper we have studied the effect of scalar leptoquarks on the rare semileptonic decays of $B$ meson. In particular, we focus
on the decay processes $B \to K l^+ l^-$ in low recoil limit and the di-neutrino decay channels $B \to K^{(*)} (X_s) \nu \bar \nu$.
The leptoquark parameter space is constrained 
by considering  the recently measured branching ratios of $B_s \to \mu ^+ \mu^-$ and $B_d \to X_s \mu^+ \mu^-$ processes.
Using the allowed parameter space we predicted the branching ratio, 
lepton non-universality factors and the  flat terms for the $B \to K l^+ l^-$ process in the low recoil region.
We found that the measured branching ratio  can be accommodated in the
 scalar leptoquark 
model. We have also calculated  the branching ratios of  $B \to K^{(*)} \nu \bar{\nu} $ and $B \to X_s \nu \bar{\nu} $ processes. 
The predicted branching ratios for $B \to K^{(*)} \nu \bar{\nu} $ processes are well below the present upper limits.
The polarization of $K^*$ and transverse asymmetry for $B \to K^* \nu \bar{\nu} $ are also computed using 
the constraint leptoquark parameters. However, we found no deviation between the SM prediction and the LQ results for different 
polarization variables and the transverse asymmetry parameter.

{\bf Acknowledgments} 

We would like to thank Science and Engineering Research Board (SERB),
Government of India for financial support through grant No. SB/S2/HEP-017/2015.

\appendix

\section{$a_l$ and $c_l$ functions in $B \to K ll$ process }
The $a_l$ and $c_l$ parameters in the decay distribution of the $B \to K l^+l^-$ processes (\ref{brK}) can be expressed as
\begin{eqnarray}
\frac{a_l}{\Gamma_0\sqrt{\lambda}\beta_lf^2_+} &=&  \frac{\lambda}{4}\left(|F_A|^2+|F_V|^2\right) +2m_l \left(m^2_B - m^2_K + q^2\right)Re\left(F_P F^*_A\right)\nn\\
&+&4m_l^2m^2_B{|F_A|}^2+q^2{|F_P|}^2 , 
\end{eqnarray}
\begin{equation}
\frac{c_l}{\Gamma_0\sqrt{\lambda}\beta_lf^2_+} =  -\beta^2_l \frac{\lambda}{4} \left(|F_A|^2+|F_V|^2\right),\hspace{5.5cm}
\end{equation}
with 
\begin{equation}
\Gamma_0 = \frac{G^2_F\alpha^2|V_{tb}V^*_{ts}|^2}{2^9\pi^5m^3_B}, \hspace{2.5cm}  \beta_l = \sqrt{1-\frac{4m_l^2}{q^2}}, \hspace{2.8cm}
\end{equation}
and
\begin{equation*}
\lambda = m_B^4 + m_K^4 + q^4 -2\left(m_B^2m_K^2 + m_B^2q^2 + m_K^2q^2\right) .
\end{equation*}

\section{Loop functions }
The loop function $X(x_t)$ in Eq. (\ref{cl}), including correction ${\cal O}(\alpha_s)$ at the next-to-leading order in QCD,
 is given by \cite{misiak, buchalla}
\begin{equation}
X(x_t) = X_0(x_t)+ \frac{\alpha_s}{4\pi}X_1(x_t),\label{b1}
\end{equation}
where 
\begin{equation}
X_0(x_t) = \frac{x_t}{8} \left[-\frac{2+x_t}{1-x_t} + \frac{3x_t-6}{(1-x_t)^2} \ln x_t\right],
\end{equation}
and 
\begin{eqnarray}
X_1(x_t)& =&  -\frac{29x_t-x_t^2-4x_t^3}{3(1-x_t)^2} - 
\frac{x_t+9x_t^2-x_t^3-x_t^4}{(1-x_t)^3 } \ln x_t \nn\\ &+& \frac{8x_t+4x_t^2+x_t^3-x_t^4}{2(1-x_t)^3 } \ln^2 x_t
 -\frac{4x_t-x_t^3}{(1-x_t)^2} L_2(1-x_t) + 8x_t \frac{\partial X_0(x_t)}{\partial x_t} \ln x_\mu.\label{b3}
\end{eqnarray}
In Eqs. (\ref{b1}-\ref{b3}), the parameters used are defined as  $x_t = m_t^2/m_W^2$, $x_\mu = \mu^2/m_W^2$ with $\mu = \mathcal{O} (m_t)$ and $L_2(1-x_t) = \int_1^{x_t} dt \frac{\ln t }{1-t}$.

\section{Transversity amplitudes for $B \to K^* \nu \bar \nu$ process }

The transversality amplitudes $A_{\perp, \parallel, 0}$ for $B \to K^* \nu \bar \nu$ process are given as 
\begin{equation}
A_\perp(s_B) = 2N\sqrt{2} \lambda^{1/2} (1,\tilde{m}_{K*}^2, s_B) (C_L^\nu + C_R^\nu) \frac{V(s_B)}{(1+\tilde{m}_{K^*})}, \hspace{4.8 cm}
\end{equation}
\begin{equation}
A_\parallel (s_B)  = -2N \sqrt{2} (1+\tilde{m}_{K^*}) (C_L^\nu - C_R^\nu) A_1 (s_B), \hspace{6.2 cm}
\end{equation}
\begin{equation}
A_0 (s_B) = -\frac{N (C_L^\nu - C_R^\nu)}{\tilde{m}_{K^*}\sqrt{s_B} } \left[ (1-\tilde{m}_{K^*}^2 - s_B) (1 + \tilde{m}_{K^*}) A_1 (s_B)
 - \lambda(1,\tilde{m}_{K^*}^2, s_B) \frac{A_2 (s_B)}{1 + \tilde{m}_{K^*}} \right],
\end{equation}
with 
\begin{equation}
N = V_{tb} V_{ts}^* \left[ \frac{G_f^2 \alpha^2 m_B^3}{3 \cdot 2^{10} \pi^5 } s_B \lambda^{1/2} (1,\tilde{m}_{K^*}^2, s_B) \right]^{1/2}.
\end{equation}
The various form factors $V(s_B)$, $A_1(s_B)$, $A_2(s_B)$ associated with $B \to K^*$ transition in Eqs. (C1-C3)  are defined as
\bea
\langle K^* \left(p_{K^*}\right)|\bar{s} \gamma _\mu  P_{L,R}  b | B\left(p\right)\rangle = i\epsilon_{\mu \nu \alpha \beta} \epsilon^{\nu *} p^\alpha q^\beta 
\frac{V(s_B)}{m_B + m_{K^*}} \mp \frac{1}{2} \Bigg(  (m_B + m_{K^*}) \epsilon^*_\mu A_1(s_B) \nn\\ 
-(\epsilon^* \cdotp q)(2p-q)_\mu \frac{A_2(s_B)}{m_B + m_{K^*}} - \frac{2m_{K^*}}{s} (\epsilon^* \cdotp q) \left[ A_3(s_B) - A_0(s_B)\right] q_\mu \Bigg),
\eea
 where $q = p_{l^+} + p_{l^-}$ and $\epsilon^\mu$ is the polarization vector of $K^*$.



\begin{thebibliography}{60}
\bibitem{lhcb1}
R. Aaij et al., [LHCb Collaboration], JHEP \textbf{1406}, 133 (2014) [arXiv:1403.8044].
\bibitem{lhcb2}
R. Aaij et al., [LHCb Collaboration], Phys. Rev. Lett. \textbf{111}, 191801 (2013) [arXiv:1308.1707].

 \bibitem{lhcb7}
  R. Aaij et al., [LHCb Collaboration], JHEP \textbf{07}, 133 (2012) [arXiv:1205.3422].
 

\bibitem{lhcb3}
R. Aaij et al., [LHCb Collaboration], Phys. Rev. Lett. \textbf{113}, 151601 (2014) [arXiv:1406.6482].

\bibitem{grinstein2}
B. Grinstein and Dan Pirjol, Phys. Rev. D \textbf{70}, 114005 (2004) [arXiv:hep-ph/0404250].

\bibitem{beylich}
   M. Beylich, G. Buchalla and T. Feldmann, Eur. Phys. J. C \textbf{71}, 1635  (2011) [arXiv: hep-ph/1101.5118]; 
G. Buchalla and G. Isidori, Nucl. Phys. B \textbf{525}, 333 (1998) [arXiv: hep-ph/9801456]. 

\bibitem{mohanta2}
 S. Sahoo and R. Mohanta, Phys. Rev. D \textbf{91}, 094019 (2015) [arXiv:1501.05193]. 


\bibitem{matias1}
S. Descotes-Genon, J. Matias, M. Ramon, J. Virto, JHEP \textbf{1301}, 048 (2013) [arXiv:1207.2753].
\bibitem{jager}
 S. J\"ager, J. Martin Camalich, JHEP \textbf{05}, 043 (2013) [arXiv:1212.2263]; S.Descotes-Genon, L. Hofer, J. Matias and J. Virto, 
JHEP \textbf{1412}, 125 (2014) [arXiv:1407.8526]; S. J\"ager and J. Martin Camalich  [arXiv:1412.3183].
 \bibitem{huber}
  T. Huber, T. Hurth and E. Lunghi, Nucl. Phys. B \textbf{802}, 40 (2008) [arXiv:0712.3009].
  
\bibitem{beaujean}
  F. Beaujean, C. Bobeth and D. van Dyk, Eur. Phys. J. C \textbf{74}, 2897 (2014) [arXiv:1310.2478];
T. Hurth and F. Mahmoudi, JHEP \textbf{04}, 097 (2014) [arXiv:1312.5267];. W. Altmannshofer, S.Gori, M. Pospelov and I. Yavin, Phys. Rev. D
 \textbf{89}, 095033 (2014) [arXiv:1403.1269]; R. Gauld, F. Goertz, and U. Haisch, JHEP \textbf{01}, 069 (2014) [arXiv:1310.1082]; 
R. Gauld, F. Goertz and U. Haisch, Phys. Rev. D \textbf{89}, 015005 (2014) [arXiv:1308.1959]; A. Datta, M. Duraisamy and D. Ghosh, 
Phys. Rev. D  \textbf{89}, 071501 (2014) [arXiv:1310.1937]; A. J. Buras and J. Girrbach, JHEP \textbf{12}, 009 (2013) [arXiv:1309.2466]; 
A. J. Buras, F. De Fazio and J. Girrbach, JHEP \textbf{02}, 112 (2014) [arXiv:1311.6729]; 
S. Descotes-Genon, J. Matias and J. Virto, Phys. Rev. D \textbf{88}, 074002 (2013) [arXiv:1307.5683]; R. R. Horgan, Z. Liu, S. Meinel and 
M. Wingate, Phys. Rev. Lett. \textbf{112}, 212003 (2014) [arXiv:1310.3887]; A. J. Buras, J. Girrbach-Noe, C. Niehof and D. M. Straub, 
JHEP {\bf 02}, 184 (2015)
[arXiv:1409.4557]; A. Crivellin, arXiv: 1409.0922; D. Ghosh, M. Nardecchia and S. Renner, JHEP \textbf{12}, 131 (2014) [arXiv:1408.4097]; 
S. Biswas, D. Chowdhury, S. Han, and S. J. Lee, 
JHEP \textbf{02}, 142 (2015)
[arXiv:1409.0882]; G. Hiller and M. Schmaltz, Phys. Rev. D \textbf{90}, 054014 (2014) [arXiv:1408.1627]; 
T. Hurth, F. Mahmoudi and S. Neshatpour, 
JHEP \textbf{12},  053 (2014)
[arXiv:1410.4545]; S. L. Glashow, D. Guadagnoli, Kenneth Lane, Phys. Rev. Lett. {\bf 114}, 091801 (2014) [arXiv:1411.0565 [hep-ph]];
 D. Aristizabal Sierra, F. Staub and A. Vicente, Phys. Rev. D \textbf{92}, 015001 (2015) [arXiv:1503.06077];
S. M. Boucenna, J. W. F. Valle and A. Vicente,  Phys. Lett. B {\bf 750}, 367 (2015) [arXiv:1503.07099];
F. Mahmoudi, S. Neshatpour, J. Virto, Eur. Phys. J. C \textbf{74},  2927, (2014) [arXiv:1401.2145];
A. Crivellin, G. D'Ambrasio, J. Heeck,  Phys. Rev. Lett. \textbf{114}, 151801 (2015) [arXiv:1501.00993];
A. Crivellin, G. D'Ambrasio, J. Heeck,  Phys. Rev. D \textbf{91}, 075006 (2015) [arXiv:1503.03477];
A. Crivellin, L. Hofer, J. Matias, U. Nierste, S. Pokorski, J. Rosiek, Phys. Rev. D {\bf 92}, 054013 (2015) [arXiv:1504.07928];
D. Becirevic, S. Faster, N. Kosnik, Phys. Rev. D \textbf{92}, 014016 (2015) [arXiv:1503.09024 [hep-hp]];
X.-W. Kang, B. Kubis, C. Hanhart, Ulf-G. Meibner, Phys Rev. D \textbf{89}, 053015 (2014), [arXiv:1312.1193];
J. Lyon, R. Zwicky, [arXiv:1406:0566];  J. Gratex, M. Hopfer, R. Zwicky, [arXiv: 1506.03970];
R. Alonso, B. Grinstein and J. M. Camalich, Phys. Rev. Lett. \textbf{113},  241802 (2014)  [arXiv:1407.7044];
R. Alonso, B. Grinstein and J. M. Camalich, JHEP \textbf{10},  184 (2015) [arXiv:1505.05164];
B. Gripaios, M. Nardecchia and S. A. Renner,  	JHEP \textbf{1505}, 006 (2015)  [arXiv:1412.1791];
A. Falkowski, M. Nardecchia, Robert Ziegler, JHEP \textbf{11},  173 (2015) [arXiv:1509.01249]; 
B. Gripaios, M. Nardecchia, S. A. Renner, [arXiv:1509.05020].
 
 
\bibitem{georgi}   H. Georgi and S. L. Glashow, Phys. Rev. Lett.\textbf{32}, 438 (1974); J. C. Pati and A. Salam, Phys.Rev. D \textbf{10}, 275 (1974).
  
 \bibitem{georgi2} 
   H. Georgi, AIP Conf. Proc. \textbf{23} 575 (1975); H. Fritzsch and P. Minkowski, Annals Phys. \textbf{93}, 193 (1975); P. Langacker, Phys. Rep. 
\textbf{72}, 185 (1981).

\bibitem{schrempp} B. Schrempp and F. Shrempp, Phys. Lett. B \textbf{153}, 101 (1985). 

  \bibitem{kaplan}   D. B. Kaplan, Nucl. Phys. B \textbf{365}, 259 (1991); B. Gripaios, JHEP \textbf{1002}, 045 (2010) [arXiv:0910.1789].

  \bibitem{davidson}    S. Davidson, D. C. Bailey and B. A. Campbell, Z. Phys.
C \textbf{61}, 613 (1994), hep-ph/9309310; I.
Dorsner, S. Fajfer, J. F. Kamenik, N. Kosnik, Phys. Lett. B
\textbf{682}, 67 (2009) [arXiv:0906.5585]; S. Fajfer, N. Kosnik, Phys. Rev. D \textbf{79}, 017502 (2009) [arXiv:0810.4858];
R. Benbrik, M. Chabab, G. Faisel, [arXiv:1009.3886]; A. V. Povarov, A. D. Smirnov, [arXiv:1010.5707]; J. P Saha, B. Misra and A. Kundu, 
Phys. Rev.D \textbf{81}, 095011 (2010), [arXiv:1003.1384]; I. Dorsner, J. Drobnak, S. Fajfer, J. F. Kamenik, N. Kosnik, JHEP \textbf{11}, 002 (2011),
[arXiv: 1107.5393]; F. S. Queiroz, K.
Sinha, A. Strumia, Phys. Rev. D \textbf{91},  035006 (2015) [arXiv:1409.6301]; B. Allanach, A. Alves, F. S. Queiroz, K. Sinha,
A. Strumia, Phys. Rev. D \textbf{92},  055023 (2015) [arXiv:1501.03494];
L. Calibbi, A. Crivellin, T. Ota,  Phys. Rev. Lett.{\bf 115}, 181801 (2015) [arXiv:1506.02661];  
Ivo de M. Varzielas and G. Hiller, JHEP {\bf 1506}, 072 (2005)  [arXiv:1503.01084];
 S. Sahoo and R. Mohanta, [arXiv:1507.02070].
\bibitem{arnold}
 J. M. Arnold, B. Fornal and M. B. Wise, Phys. Rev. D \textbf{88}, 035009 (2013), [arXiv:1304.6119].

\bibitem{kosnik}
  N. Kosnik, Phys. Rev. D \textbf{86}, 055004 (2012), [arXiv:1206.2970].

 
 \bibitem{mohanta1}
 R. Mohanta, Phys. Rev. D \textbf{89}, 014020 (2014) [arXiv:1310.0713].

\bibitem{leptoquark}
D. Aristizabal Sierra, M. Hirsch, S. G. Kovalenko, Phys. Rev. D \textbf{77}, 055011 (2008), [arXiv:0710.5699]; 
K. S. Babu, J. Julio, Nucl. Phys. B \textbf{841}, 130 (2010), [arXiv:1006.1092]; S. Davidson, S. Descotes-Genon, JHEP \textbf{1011}, 073 (2010), 
[arXiv:1009.1998]; S. Fajfer, J. F. Kamenik, I. Nisandzic, J. Zupan, Phys. Rev. Lett. \textbf{109}, 161801, (2012), [arXiv:1206.1872]; 
K. Cheung, W.-Y. Keung, P.-Y. Tseng, [arXiv:1508.01897]; D. A. Camargo, [arXiv:1509.04263]; S. Baek, K. Nishiwaki, [arXiv:1509.07410].


\bibitem{buras1} C. Bobeth, M. Misiak and J. Urban, Nucl. Phys. B {\bf 574}, 291
(2000) [hep-ph/9910220];
C. Bobeth, A. J. Buras, F. Kr\"uger and J. Urban, Nucl. Phys. B {\bf 630}, 87 (2002) [hep-ph/0112305].

\bibitem{kohda}
W.-S. Hou, M. Kohda and F. Xu, Phys. Rev. D \textbf{90}, 013002 (2014) [arXiv:1403.7410].

\bibitem{ref45} W. Buchmuller, R. Ruckl and D. Wyler, Phys. Lett. B {\bf 191}, 442 (1987); Erratum-{\it ibid.}
B {\bf 448}, 320 (1997). 
 
 
 \bibitem{bobeth1}
C. Bobeth, M. Gorbahn, T. Hermann, M. Misiak, E. Stamou, M. Steinhauser, Phys. Rev. Lett.
\textbf{112}, 101801 (2014) [arXiv:1311.0903].
\bibitem{cms}
S. Chatrchyan et al., [CMS Collaboration], Phys. Rev. Lett.\textbf{111}, 101805 (2013),
[arXiv:1307.5025].
\bibitem{lhcb5}
 R. Aaij et al., [LHCb Collaboration], Phys. Rev. Lett. \textbf{111}, 101805 (2013), [arXiv:1307.5024].
\bibitem{lhcb6}
CMS and LHCb Collaborations, EPS-HEP 2013 European Physical Society Conference on High Energy Physics, Stockholm, Sweden, 2013, 
Conference Report No. CMS-PAS-BPH-13-007, LHCb-CONF-2013-012, http://cds.cern.ch/record/1564324.


 
 
 \bibitem{bobeth2}
C. Bobeth, G. Hiller, D. van Dyk and C. Wacker, \textbf{JHEP 1201}, 107 (2012)  [arXiv:1111.2558].

\bibitem{bobeth4}
C. Bobeth, G. Hiller and D. van Dyk, JHEP \textbf{1007}, 098 (2010) [arXiv:1006.5013].


\bibitem{grinstein} 
 B. Grinstein and D. Pirjol, Phys. Lett. B \textbf{533}, 8 (2002)  [arXiv:hep-ph/0201298].



 \bibitem{bobeth3}
C. Bobeth, G. Hiller and G. Piranishvili, JHEP \textbf{0712}, 040 (2007) [arXiv:0709.4174].

\bibitem{bobeth5}
C. Bobeth, G. Hiller and D. van Dyk, JHEP \textbf{1107}, 067 (2011) [arXiv:1105.0376].

\bibitem{mannel}
  A. Khodjamirian, T. Mannel, A. A. Pivovarov and Y. M. Wang, JHEP \textbf{1009}, 089 (2010) [arXiv:1006.4945].


\bibitem{pdg}
 K.A. Olive et al. (Particle Data Group), Chin. Phys. C {\bf 38}, 090001 (2014).

\bibitem{buras2}
W. Altmannshofer, A. J. Buras, D.M. Straub and M. Wick, JHEP \textbf{0904}, 022 (2009) [arXiv:0902.0160].

\bibitem{fazio}
 P. Colangelo, F. De Fazio, P. Santorelli, and E. Scrimieri, Phys. Lett. B {\bf 395}, 339 (1997)  [arXiv: hep-ph/9610297].

 

\bibitem{ball3}
 P. Ball and R. Zwicky, Phys. Rev.D \textbf{71}, 014015 (2005) [arXiv: hep-ph/0406232].


\bibitem{kim}
 C. S. Kim, Y. G. Kim and T. Morozumi, Phys.Rev.D \textbf{60},  094007 (1999) [arXiv: hep-ph/9905528].


 \bibitem{egede}
U. Egede, T. Hurth, J. Matias, M. Ramon, W. Reece, JHEP \textbf{11}, 032 (2008) [arXiv: 0807.2589].


 
 


 \bibitem{simula}
  D. Melikhov, N. Nikitin and S. Simula, Phys. Lett. B \textbf{428}, 171 (1998)  [arXiv: hep-ph/9803269].


\bibitem{ball1}
W. Altmannshofer, P. Ball, A. Barucha, A. J. Buras, D. M. Straub and M. Wick, JHEP \textbf{01}, 019 (2009) [arXiv:0811.1214].


\bibitem{ball2}
P. Ball and R. Zwicky, Phys. Rev. D \textbf{71}, 014029 (2005) [arXiv: hep-ph/0412079].


 
\bibitem{nardi}
 Y. Grossman, Z. Ligeti and E. Nardi, Nucl. Phys.\textbf{B 465}, 369 (1996) [arXiv: hep-ph/9510378];
 C. Bobeth, A. J. Buras, F. Kruger and J. Urban, Nucl. Phys. \textbf{B 630}, 87 (2002)  [arXiv: hep-ph/0112305];  
G. Buchalla, A.J. Buras and M. E. Lautenbacher, Rev. Mod. Phys.\textbf{ 68}, 1125 (1996)  [arXiv: hep-ph/9512380].

 

\bibitem{misiak}
 M. Misiak and J. Urban, Phys. Lett. \textbf{B 451}, 161 (1999)  [arXiv: hep-ph/9901278].
 \bibitem{buchalla}
  G. Buchalla and A. J. Buras, Nucl. Phys. \textbf{B 548}, 309 (1999)  [arXiv:hep-ph/9901288].


 
\end{thebibliography}
\end{document}